\documentclass[jgrga,arxiv,times]{agutex4}
\usepackage{amsmath,amssymb}
\usepackage{graphicx,color,soul}
\usepackage{lineno}
\usepackage{mathptmx}

\setkeys{Gin}{draft=false}

\authorrunninghead{BRANTUT ET AL.}
\titlerunninghead{SLIP PULSE STABIILITY}

\authoraddr{N. Brantut, Department of Earth Sciences, University College London, Gower Street, London WC1E 6BT, UK. (n.brantut@ucl.ac.uk)}

\begin{document}

\title{Stability of pulse-like earthquake ruptures}

\authors{Nicolas Brantut\altaffilmark{1}, Dmitry I. Garagash\altaffilmark{2}, Hiroyuki Noda\altaffilmark{3}}

\altaffiltext{1}{Department of Earth Sciences, University College London, London, UK.}
\altaffiltext{2}{Department of Civil and Resource Engineering, Dalhousie University, Halifax, Canada.}
\altaffiltext{3}{Disaster Prevention Research Institute, Kyoto University, Uji, Japan.}

\begin{abstract}
  Pulse-like ruptures arise spontaneously in many elastodynamic rupture simulations and seem to be the dominant rupture mode along crustal faults. Pulse-like ruptures propagating under steady-state conditions can be efficiently analysed theoretically, but it remains unclear how they can arise and how they evolve if perturbed. Using thermal pressurisation as a representative constitutive law, we conduct elastodynamic simulations of pulse-like ruptures and determine the spatio-temporal evolution of slip, slip rate and pulse width perturbations induced by infinitesimal perturbations in background stress. These simulations indicate that steady-state pulses driven by thermal pressurisation are unstable. If the initial stress perturbation is negative, ruptures stop; conversely, if the perturbation is positive, ruptures grow and transition to either self-similar pulses (at low background stress) or expanding cracks (at elevated background stress). Based on a dynamic dislocation model, we develop an elastodynamic equation of motion for slip pulses, and demonstrate that steady-state slip pulses are unstable if their accrued slip $b$ is a decreasing function of the uniform background stress $\tau_\mathrm{b}$. This condition is satisfied by slip pulses driven by thermal pressurisation. The equation of motion also predicts quantitatively the growth rate of perturbations, and provides a generic tool to analyse the propagation of slip pulses. The unstable character of steady-state slip pulses implies that this rupture mode is a key one determining the minimum stress conditions for sustainable ruptures along faults, i.e., their ``strength''. Furthermore, slip pulse instabilities can produce a remarkable complexity of rupture dynamics, even under uniform background stress conditions and material properties.
\end{abstract}

\begin{article}

  \section{Introduction}

  The propagation of earthquakes is generally classified into two main modes: crack-like ruptures, where fault slip occurs throughout the duration of propagation, and pulse-like ruptures, where only a small portion of the fault inside a ruptured area is sliding at a given time during rupture. The observation that local slip duration is often much shorter than the time required for stopping phases to propagate from fault boundaries led \citet{heaton90} to suggest that most crustal earthquakes may propagate as pulse-like ruptures. A number of detailed kinematic and dynamic inversions of earthquake slip \citep{wald94,beroza96,olsen97,day98,galetzka15} have confirmed the pulse-like nature of large crustal earthquakes, highlighting the importance of this rupture mode in the physics of faults.

  The physical origin and dynamics of pulse-like ruptures have been studied extensively in theoretical models. Slip events have been shown to propagate as narrow, self-similar slip pulses in simplified discrete spring block models \citep[e.g.,][]{carlson89, elbanna12}. Fully dynamic rupture simulations have revealed the key role of velocity-weakening friction \citep[e.g.][]{heaton90,cochard94,perrin95,zheng98} and boundary conditions \citep[e.g.][]{johnson90,johnson92e} in the spontaneous generation of slip pulses. Specifically, elastodynamic simulations with velocity-dependent friction show that the existence and evolution of the dynamic pulse-like ruptures  are strongly controlled by both the ambient background stress and the nucleation conditions on the fault \citep{zheng98,gabriel12}. For a given nucleation condition, an increase in background stress results in a sequential transition from arresting pulses to growing pulses, and then growing crack-like ruptures. Therefore, the mode of rupture and its evolution are the signature of the background stress acting on the fault prior to the earthquake. Of critical importance here is the stress level at the transition from arresting to growing pulses, which provides the threshold below which sustained fault slip is precluded. The rupture mode at the transition is that of a ``steady-state'' slip pulse, for which the tip and tail of the slipping patch propagate at the same speed. These steady-state solutions are therefore key to understand the stress level required for earthquake propagation and the dynamics of faults.

  Steady-state solutions of the elastodynamic fault problem can be obtained using analytical or simple numerical methods, so that they can be studied efficiently without resorting to computationally expensive numerical treatment. Several pulse-like rupture solutions have been obtained for simple models of faults with a constant or slip-dependent friction law \citep[e.g.][]{broberg78,freund79,rice05,dunham05}, but without specific regards to the processes allowing for strength recovery and ``healing'' (i.e., cessation of slip) at the tail of the pulse. Steady-state pulse solutions fully consistent with both elastodynamics and a specific friction law have been determined by \citet{perrin95} in the context of rate-and-state friction, and more recently by \citet{garagash12} and \citet{platt15b} in the context of dynamic weakening by thermal (or chemical) pressurisation of pore fluids within the fault zone. These solutions provide unique insights into the relationships between rupture properties, such as pulse width or rupture velocity at a given background stress, and key parameters of the friction law, such as rate-and-state parameters \citep{perrin95} or thermo-hydraulic properties of the fault core \citep{garagash12, platt15b}. Despite the (relative) simplicity and efficiency of those steady state pulse solutions, it remains to be confirmed how they can be generated and how they evolve in response to perturbations in loading conditions or frictional properties. In other words, the key question here is to determine how self-consistent steady-state solutions (i.e., satisfying elastodynamics and all the features of a specific friction law) can be compared to possibly transient rupture dynamics observed on natural faults.

  
  Regarding this issue, the numerical simulations provided by \citet{gabriel12} and \citet{brener18} using velocity-dependent friction, or by \citet{noda09} in the context of weakening by thermal pressurisation of pore fluids within the fault, seem to indicate that such steady-state solutions are not stable: they either grow (to form self-similar pulses) or decay and stop. The goal of this paper is to analyse in detail how steady-state pulses respond to perturbations and to determine a clear stability condition depending on the characteristics of the friction law. Building on the work by \citet{garagash12}, we examine specifically the case of pulses driven by thermal pressurisation of pore fluids, and first solve the nonlinear perturbation problem numerically (Section \ref{sec:TPnum}). We then examine more generally the conditions under which stable pulses can exist based on an approximate equation of motion for moving dislocations (Section \ref{sec:EoM}). The significance of steady-state pulse solutions and some implications for the dynamics of earthquakes are examined in Section \ref{sec:discussion}.


\section{Slip pulses driven by thermal pressurisation of pore fluids}
\label{sec:TPnum}

In this Section, we present a detailed analysis of the evolution of pulses driven by thermal pressurisation. We choose to focus specifically on thermal pressurisation as the governing process by which faults weaken (and restrengthen), since it has a firm physical background, and has been shown to be consistent with a number of seismological observations \citep{rice06,viesca15}. Beyond this specific choice for the fault constitutive behaviour, we stress that the method of analysis developed here is quite general and can be used to include other friction laws.

We first briefly summarise the results of \citet{garagash12} regarding steady-state solutions, and perform a stability analysis by solving for the evolution of perturbations from the steady-state solution.

\begin{figure}
  \centering
  \includegraphics{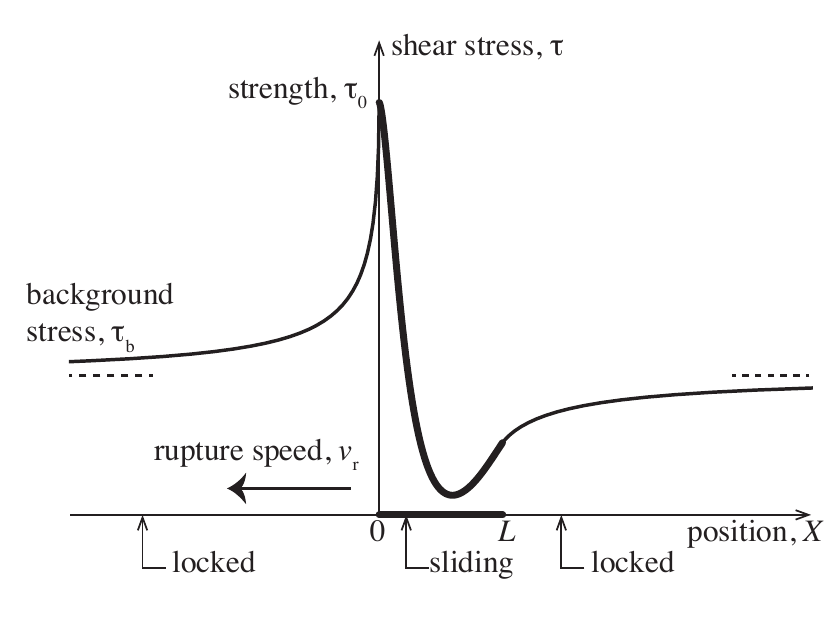}
  \caption{Schematic of the stress evolution along a pulse-like rupture shown in the coordinate frame $X$ moving with the rupture front. Far from the slipping patch, the shear stress is constant and equal to $\tau_\mathrm{b}$. Near the rupture tip ($X=0$), the stress increases up to the local strength $\tau_0$, and then evolves according to a constitutive law, in agreement with elastodynamic equilibrium. Behind the patch ($X=L$), the stress increases again back to the background stress.}
  \label{fig:pulsesketch}
\end{figure}

\subsection{Model and steady-state solution}
\label{sec:ss}

\subsubsection{Elastodynamics of steadily-propagating pulse}

We consider a planar fault embedded in an infinite, homogeneous elastic medium of shear modulus $\mu$. The fault is assumed to be of infinite extent in one of its planar dimensions, so that we restrict our attention to a two-dimensional problem. The fault is loaded by a uniform background shear stress, denoted $\tau_\mathrm{b}$. For simplicity, we assume that the loading is in mode III (out of plane) geometry. Fault slip is assumed to occur over a patch of finite length $L$, which propagates at a constant speed $v_\mathrm{r}$ along spatial coordinate $x$, as shown in Figure \ref{fig:pulsesketch}. Under steady-state conditions (i.e., constant rupture speed), it is convenient to introduce a reference frame $(X,y)$ that moves with the rupture tip, so that the shear stress $\tau$ and slip rate $V$ along the fault (in the plane $y=0$) are functions of the coordinate $X = v_\mathrm{r}t - x$ only. The elastodynamic equilibrium requires that \citep[e.g.][]{weertman69}
\begin{linenomath}
  \begin{equation} \label{eq:pulse_eq}
    \tau(X) = \tau_\mathrm{b} - \frac{\bar{\mu}}{2\pi v_\mathrm{r}}\int_0^L\frac{V(\xi)}{X-\xi}d\xi,
  \end{equation}
\end{linenomath}
where $\bar\mu$ is an apparent shear modulus given by $\bar\mu = \mu\times F(v_\mathrm{r}/c_\mathrm{s})$. The function $F$ of the ratio of rupture speed and shear wave speed $c_\mathrm{s}$ is equal to $F(v_\mathrm{r}/c_\mathrm{s}) = \sqrt{1 - v_\mathrm{r}^2/c_\mathrm{s}^2}$ (\citep[e.g.][]{rice80}), so that the apparent modulus approaches zero as the rupture speed approaches the shear wave speed. In Equation \eqref{eq:pulse_eq}, it is understood that the slip rate is given by
\begin{linenomath}
  \begin{equation}
    V(X) = v_\mathrm{r}\frac{d\delta}{dX},
  \end{equation}
\end{linenomath}
where $\delta$ is the slip.

In the slipping part of the fault ($0\leq X\leq L$), the stress $\tau(X)$ must be equal to the fault strength $\tau_\mathrm{f}$, which is given by a constitutive law (see below). Furthermore, at the tail of the pulse ($X\geq L$), we need to ensure that the strength remains higher than the elastic stress $\tau(X)$ (otherwise slip would continue, which would be in contradiction with the pulse width being equal to $L$). \citet{garagash12} determined that the stress gradient at the tail of the pulse is singular, of the form $d\tau/dX \propto k_\mathrm{L}/\sqrt{X-L}$, where
\begin{linenomath}
  \begin{equation} \label{eq:kL}
    k_\mathrm{L} = -\frac{4}{\pi\sqrt{L}}\int_0^L\sqrt{\frac{X}{L-X}}\frac{d\tau}{dX}dX.
  \end{equation}
\end{linenomath}
The gradient in fault strength remains continuous, so that the condition for cessation of slip $\tau(X)\leq\tau_\mathrm{f}$ imposes that the elastic stress gradient remains bounded, i.e., $k_\mathrm{L}=0$. This equality ensures the consistency of the assumption that slip only occurs where $\tau_\mathrm{f}=\tau(X)$, and provides a constraint on the pulse length $L$ (which would otherwise be a free parameter of the problem).

\begin{figure}
  \centering
  \includegraphics{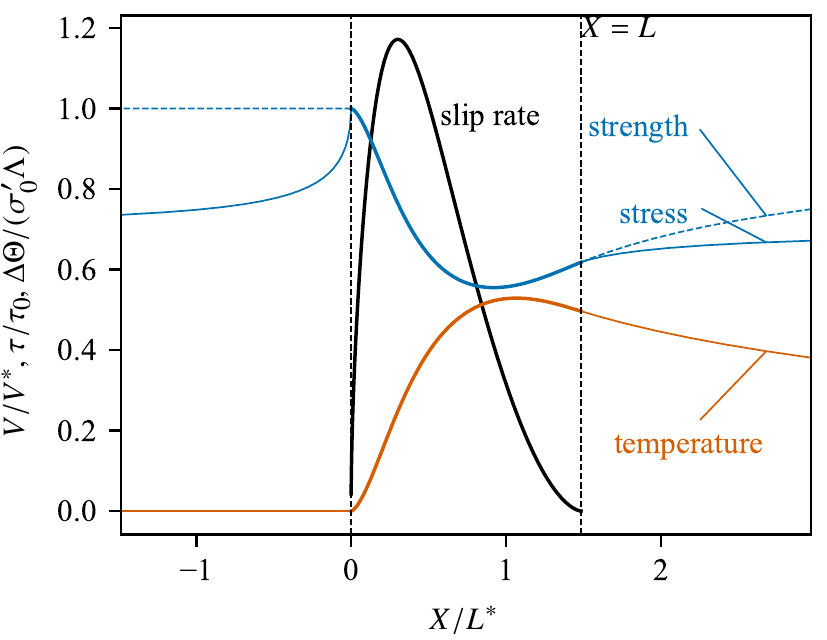}
  \caption{Example solution of a steady-state pulse driven by thermal pressurisation. Vertical dashed lines indicate the beginning and end of the slipping patch. The background stress is $\tau_\mathrm{b}=0.7$ and the diffusivity ratio is $\alpha_\mathrm{hy}/\alpha_\mathrm{th}=1$. Using $h/h_\mathrm{dyna}=1$, the resulting pulse speed is $v_\mathrm{r}/c_\mathrm{s}=0.894$, length is $L/L^*=1.485$, duration is $T/T^*=1.661$ and total slip is $b/\delta_\mathrm{c}=0.974$.}
  \label{fig:pulseexample}
\end{figure}

\subsubsection{Fault strength}

\begin{figure*}
  \centering
  \includegraphics{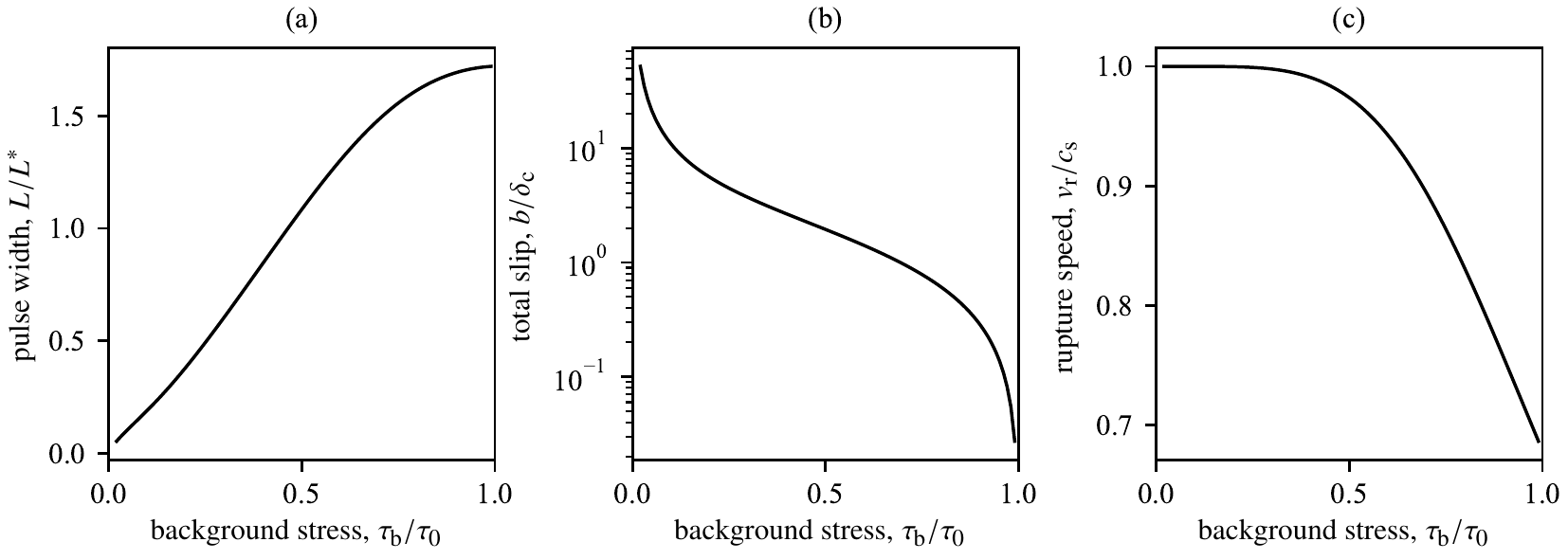}
  \caption{Pulse width (a), total slip (b) and rupture speed (c) as functions of the background stress for steady-state pulses driven by thermal pressurisation, assuming $\alpha_\mathrm{hy}/\alpha_\mathrm{th}=1$.}
  \label{fig:pslip_taub}
\end{figure*}

The strength of the fault $\tau_\mathrm{f}$ is assumed to be governed by a friction law:
\begin{linenomath}
  \begin{equation}\label{eq:tauf}
    \tau_\mathrm{f} = f\sigma' = f\times(\sigma_\mathrm{n}-p),
  \end{equation}
\end{linenomath}
where $f$ is a friction coefficient and $\sigma'$ is the Terzaghi effective stress, equal to the difference between the fault normal stress $\sigma_\mathrm{n}$ and the pore fluid pressure $p$ inside the fault core. Here, we assume a constant friction coefficient throughout the slip process. Since we are primarily concerned here with dynamic slip, the constant value of $f$ should be representative of the high velocity ``dry'' friction coefficient, which is typically of the order of $0.1$ \citep[e.g.][]{ditoro11}. The pore fluid pressure $p$ is governed by the competition between fluid diffusion and thermal expansion due to shear heating. The fluid pressure evolution is coupled to the temperature $\Theta$ through the following Equations \citep[e.g.][]{rice06}:
\begin{linenomath}
  \begin{align}
    \frac{\partial p}{\partial t} &= \Lambda \frac{\partial \Theta}{\partial t} + \alpha_\mathrm{hy}\frac{\partial^2p}{\partial y^2}, \label{eq:dpdt}\\
    \frac{\partial \Theta}{\partial t} &= \frac{\tau_\mathrm{f}\dot\gamma}{\rho c} + \alpha_\mathrm{th}\frac{\partial^2 \Theta}{\partial y^2}, \label{eq:dTdt}
  \end{align}
\end{linenomath}
where $\Lambda$ is a thermo-poro-elastic coupling factor expressing the increase in fluid pressure per unit increase in temperature, $\alpha_\mathrm{hy}$ and $\alpha_\mathrm{th}$ are hydraulic and thermal diffusivities, respectively, $\dot\gamma$ is the shear strain rate, and $\rho c$ is the heat capacity of the fault rock. In a fault core of finite width, pore pressure is not homogeneous across the fault, which can lead to shear strain localisation \citep{rice14,platt14}. Here, we do not explicitly account for this effect, which requires the introduction of further parameters such as rate-hardening properties of the sheared gouge. We follow \citet{garagash12} and consider a Gaussian shear strain rate distribution across the fault with a characteristic width $h$,
\begin{linenomath}
  \begin{equation} \label{eq:gdot}
    \dot\gamma(y,t) = \frac{V(t)}{h}e^{-\pi y^2/h^2},
  \end{equation}
\end{linenomath}
and use the pore pressure at the center of the fault (where it is maximum) to compute the strength in Equation \eqref{eq:tauf}. 

Equations \eqref{eq:dpdt} and \eqref{eq:dTdt} can be solved to arrive at the integral representation for fault strength \citep{rice06} given here in the form  of \citep{garagash12}:
\begin{linenomath}
  \begin{equation} \label{eq:tauf_integral}
    \tau_\mathrm{f} = \tau_0 - \frac{1}{\delta_\mathrm{c}}\int_0^t\tau_\mathrm{f}(t')V(t')\mathcal{K}\left(\frac{t-t'}{T^*}; \frac{\alpha_\mathrm{hy}}{\alpha_\mathrm{th}}\right)dt',
  \end{equation}
\end{linenomath}
where $\tau_0 = f\sigma'_0$ is the initial strength of the fault (at $p(y=0,t=0)=p_0$), and
\begin{linenomath}
  \begin{equation}\label{eq:dc_Tstar}
    \delta_\mathrm{c}=\frac{\rho c}{f\Lambda}h \quad\text{and}\quad T^* = \frac{h^2}{4\alpha},
  \end{equation}
\end{linenomath}
where $\alpha = (\sqrt{\alpha_\mathrm{th}} + \sqrt{\alpha_\mathrm{hy}})^2$, are characteristic slip weakening distance and diffusion time, respectively. The convolution kernel $\mathcal{K}$ is given in Appendix \ref{ax:convolution}.

\subsubsection{Steady-state pulse solution}

For a given background stress $\tau_\mathrm{b}/\tau_0$ and diffusivity ratio $\alpha_\mathrm{hy}/\alpha_\mathrm{th}$, Equations \eqref{eq:pulse_eq} and \eqref{eq:tauf_integral}, under the condition \eqref{eq:kL}, have been solved for slip rate $V$ and strength $\tau_\mathrm{f}$ by \citet{garagash12}. Here, we reproduce these computations using a more efficient quadrature method given by \citet{viesca18} (see Appendix \ref{ax:GC} for more details). To keep the solutions as general as possible, we normalise the stresses by $\tau_0$, slip by $\delta_\mathrm{c}$, time by $T^*$, and slip rate and distance by
\begin{linenomath}
  \begin{equation}
    V^* = \delta_\mathrm{c}/T^* \quad\text{and}\quad L^* = \mu\delta_\mathrm{c}/\tau_0.
  \end{equation}
\end{linenomath}
In the determination of the solution, we constrain not only the distribution of stress and slip rate along the pulse, but also its duration $T/T^*$ and length $L/L^*$. The rupture speed is given by $v_\mathrm{r}=L/T$, so that its ratio relative to the shear wave speed is $v_\mathrm{r}/c_\mathrm{s}=(L/L^*)(T^*/T)/(c_\mathrm{s}T^*/L^*)$. Following \citet[Section 7.1]{garagash12}, we define a characteristic thickness $h_\mathrm{dyna}$ such a that $h/h_\mathrm{dyna} = c_\mathrm{s}T^*/L^*$, i.e., 
\begin{linenomath}
  \begin{equation}
    h_\mathrm{dyna} = \frac{\mu}{\tau_0}\frac{\rho c}{f \Lambda}\frac{4\alpha}{c_\mathrm{s}},
  \end{equation}
\end{linenomath}
so that constraining the fault core thickness through the ratio $h/h_\mathrm{dyna}$ implies that the rupture velocity $v_\mathrm{r}/c_\mathrm{s}$ is also constrained ($v_\mathrm{r}/c_\mathrm{s} = (L/L^*)(T^*/T)(h_\mathrm{dyna}/h)$).

A representative example is shown in Figure \ref{fig:pulseexample}, where we chose $\tau_\mathrm{b}/\tau_0=0.7$ and $\alpha_\mathrm{hy}/\alpha_\mathrm{th}=1$. For completeness, we also show the evolution of stress $\tau$ and strength $\tau_\mathrm{f}$ outside the pulse, and we indeed observe that $\tau<\tau_\mathrm{f}$ behind the tail. This pulse is therefore fully consistent with elastodynamics and the fault constitutive law.

Some key properties of steady-state pulses driven by thermal pressurisation can be determined from a systematic exploration of the numerical solutions. Of particular interest here are the pulse width ($L/L^*$), total slip ($b/\delta_\mathrm{c}$) and rupture speed ($v_\mathrm{r}/c_\mathrm{s}$), which are shown in Figure \ref{fig:pslip_taub} as a function of the background stress ($\tau_\mathrm{b}/\tau_0$). In all these plots, we chose again $\alpha_\mathrm{hy}/\alpha_\mathrm{th}=1$, knowing that this parameter has only a minor quantitative effect on the results \citep{garagash12}.

With increasing background stress, the slip and rupture speed decrease, while the pulse width increases. However, the relationships depicted in Figure \ref{fig:pslip_taub} have been derived from independent steady-state solutions, and therefore may not correspond to the actual \emph{evolution} of the width, slip and rupture speed of a \emph{single} pulse propagating along a fault with varying background stresses, pore pressure, or fault constitutive parameters (friction, etc). In order to compute such an evolution, and determine whether a given steady-state solution is stable against perturbations in background stress, we need to compute the full elastodynamic solution for a propagating pulse in a perturbed stress state.



\subsection{Elastodynamic stability analysis: Method}


The elastodynamic stress equilibrium can be expressed as
\begin{linenomath}
  \begin{equation} \label{eq:elastodynamics_full}
    \tau(x,t) = \tau_\mathrm{b}(x) - \frac{\mu}{2c_\mathrm{s}} V(x,t) + \phi[V],
  \end{equation}  
\end{linenomath}
where $\phi[V]$ is a linear functional of slip rate that corresponds to the stress redistribution due to slip and elastic waves. In Equation \eqref{eq:elastodynamics_full}, an explicit space dependency has been written for the background stress, $\tau_\mathrm{b}(x)$, to account for the introduction of local perturbations. Direct solutions of Equation \eqref{eq:elastodynamics_full} can be obtained numerically, but require somewhat arbitrary rupture initiation conditions, which would be incompatible with our objective of studying small perturbations around a steadily propagating rupture, regardless of how this rupture originated. Here, we circumvent the rupture nucleation problem and only solve for stress and slip rate perturbations from a preexisting steady-state pulse solution.

Let us denote $\tau_\mathrm{ss}(x,t)$, $\delta_\mathrm{ss}(x,t)$ and $V_\mathrm{ss}(x,t)$ the stress, slip and slip rate associated with a steady-state pulse propagating along a fault under a uniform background stress $\tau_\mathrm{b,ss}$. By construction, $\tau_\mathrm{ss}$, $\delta_\mathrm{ss}$ and $V_\mathrm{ss}$ are solutions of Equation \eqref{eq:elastodynamics_full} with $\tau_\mathrm{b}=\tau_\mathrm{b,ss}$. Now if we introduce a perturbation in background stress $\Delta\tau_\mathrm{b}(x)$, the resulting perturbations $\Delta\tau(x,t)$, $\Delta\delta(x,t)$ and $\Delta V(x,t)$ in stress, slip and slip rate, respectively, satisfy
\begin{linenomath}
  \begin{equation} \label{eq:elast_pert}
    \Delta\tau(x,t) = \Delta\tau_\mathrm{b}(x) - \frac{\mu}{2c_\mathrm{s}}\Delta V(x,t) + \phi[\Delta V],
  \end{equation}
\end{linenomath}
where we made use of the linearity of the functional $\phi$. The strength evolution due to thermal pressurisation is given in Equation \eqref{eq:tauf_integral}, which is rewritten in terms of strength and slip rate perturbations as
\begin{linenomath}
  \begin{align} 
    \Delta\tau_\mathrm{f}(x,t) &= -\frac{1}{\delta_\mathrm{c}}\int_0^t\left(\tau_\mathrm{f,ss}(x,t')\Delta V(x,t')+\Delta\tau_\mathrm{f}(x,t')V_\mathrm{ss}(x,t')\right.\nonumber\\
    &\quad \left.+\Delta\tau_\mathrm{f}(x,t')\Delta V(x,t')\right)\mathcal{K}\left(\frac{t-t'}{T^*};\frac{\alpha_\mathrm{hy}}{\alpha_\mathrm{th}}\right)dt', \label{eq:dtauf}
  \end{align}
\end{linenomath}
where $\tau_\mathrm{f,ss}(x,t)$ is the strength along the steady-state pulse and $\Delta\tau_\mathrm{f}$ is the strength perturbation. The governing Equation \eqref{eq:dtauf} for the strength perturbation is not linear, and therefore requires the specific knowledge of the steady-state strength and slip rate profiles, $\tau_\mathrm{f,ss}(x,t)$ and $V_\mathrm{ss}(x,t)$. These profiles correspond to the solutions of the steady-state problem stated in the previous Section.

\begin{figure*}[b]
  \centering
  \includegraphics{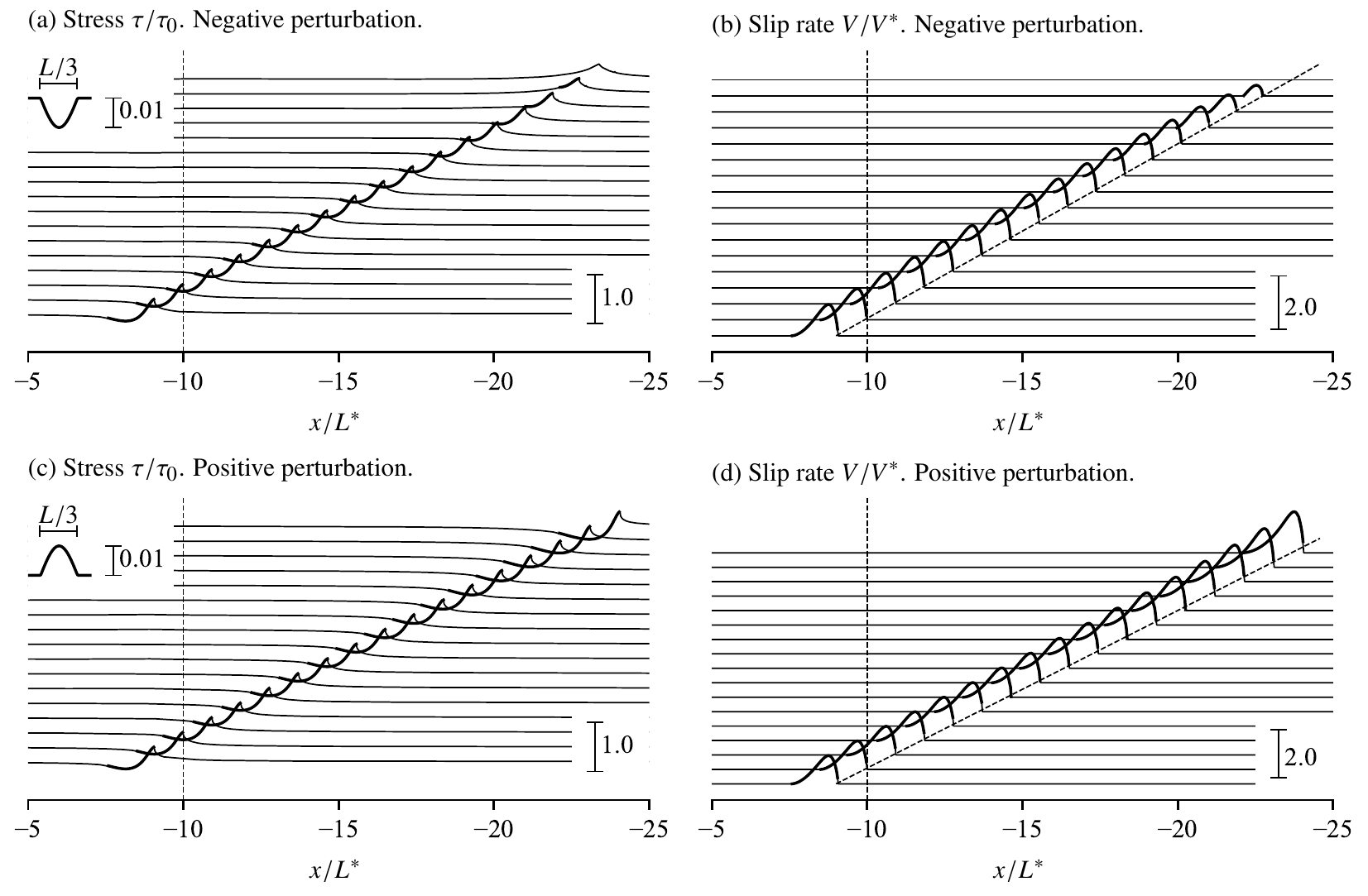}
  \caption{Snapshots of shear stress and slip rate profiles for a slip pulse propagating from left to right across a negative (a,b) or positive (c,d) background stress perturbation. The initial steady state pulse is generated with a background stress $\tau_\mathrm{b}/\tau_0=0.7$, diffusivity ratio $\alpha_\mathrm{hy}/\alpha_\mathrm{th}=1$ and $h/h_\mathrm{dyna}=1$. The perturbation amplitude is $|\Delta\tau_\mathrm{b}|/\tau_0=10^{-2}$, with a half-sinusoidal shape of width $L/3$ (see insets in panels a and c), centered at $x/L^*=-10$ (dotted line). In all the plots, thick lines mark the positions where slip rate is nonzero. Snapshots are shown at regular time intervals of $\approx 1.03T^*$. Oblique dashed lines in panels (b) and (d) show the virtual position of the steady-state rupture tip without perturbation (i.e., rupture speed $v_\mathrm{r,ss}$).}
  \label{fig:pertexample1}
\end{figure*}

Our solution strategy therefore consists in first solving a steady-state problem (see previous Section), and then solving the full elastodynamic problem for perturbations to this solution arising from variations in background stress. In practice, we use the spectral boundary integral method of \citet{perrin95,lapusta00,noda10} to compute the dynamic stress distribution functional $\phi[\Delta V]$ (see Appendix \ref{ax:func}), and a predictor-corrector method for time integration. The details of the algorithm are given in Appendix \ref{ax:numerics}.

\subsection{Elastodynamic stability analysis: Results}

\begin{figure*}
  \centering
  \includegraphics{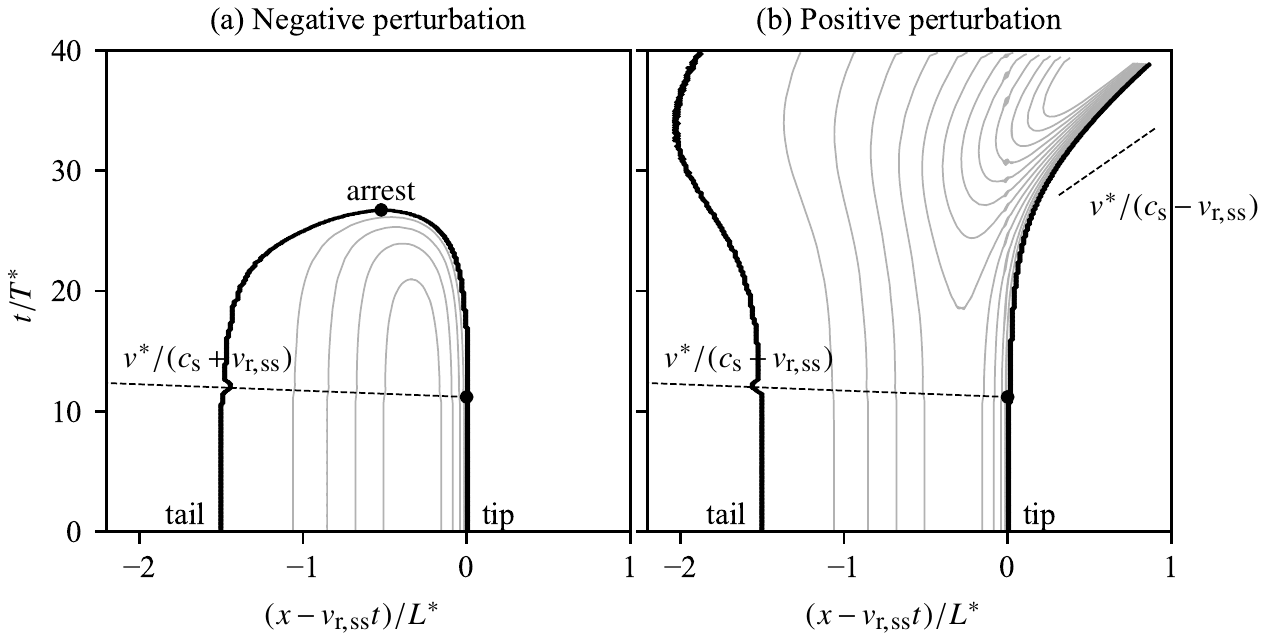}
  \caption{Slip pulse shape in transformed coordinates $\big((x-v_\mathrm{r,ss}t)/L^*, t/T^*\big)$ for a negative (a) and positive (b) perturbation. The initial background stress is $\tau_\mathrm{b}/\tau_0=0.7$, and the perturbation amplitude is $|\Delta\tau_\mathrm{b}|/\tau_0=10^{-2}$. Black contours mark where slip rate is zero (i.e., delimit the pulse tip and tail positions). Grey lines are slip rate contours, spaced by $0.25 V^*$ increments. Black dot marks the position of the perturbation, and dotted lines highlight the shear wave fronts emitted from the perturbation. The normalising speed is given by $v^*=L^*/T^*$.}
  \label{fig:pulseshape}
\end{figure*}

Two representative examples of slip pulse propagating across either a positive or a negative perturbation in background stress are shown in Figure \ref{fig:pertexample1}. The initial background stress is $\tau_\mathrm{b}/\tau_0=0.7$ and the diffusivity ratio is $1$. In both cases, the perturbation was a half-sine of $(1/3)L/L^*$ in width and $\Delta\tau_\mathrm{b}/\tau_0=10^{-2}$ in amplitude. When crossing a negative perturbation (Figure \ref{fig:pertexample1}a,b), the slip pulse continues to propagate over a distance of the order of $10 L^*$ while both the dynamic stress drop and slip rate progressively reduce, until rupture arrests. Conversely, a positive perturbation (Figure \ref{fig:pertexample1}c,d) amplifies the dynamic stress drop and slip rate, and also results in a progressive increase in pulse width.

\begin{figure*}
  \centering
  \includegraphics{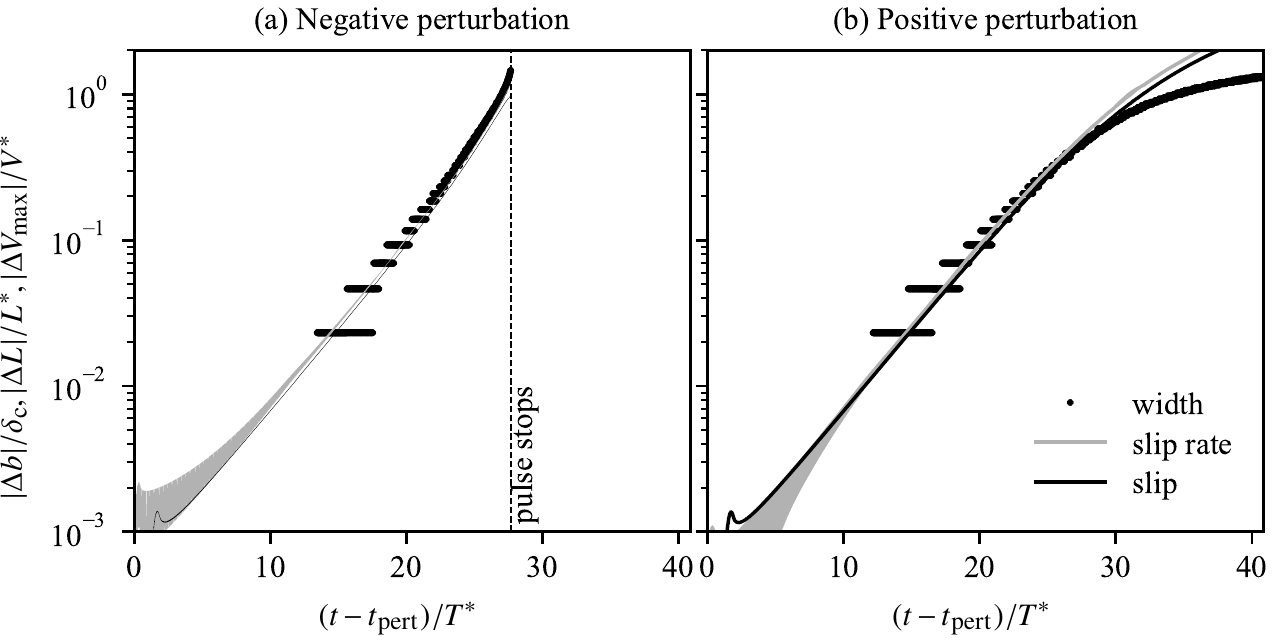}
  \caption{Time evolution of perturbations in pulse width ($|\Delta L|/L^*$, black dots), peak slip rate ($|\Delta V_\mathrm{max}|/V^*$, grey lines) and net slip ($|\Delta b|/\delta_\mathrm{c}$, black lines) in the case of a negative (a) and positive (b) perturbation. The initial background stress is $\tau_\mathrm{b}/\tau_0=0.7$, and the perturbation amplitude is $|\Delta\tau_\mathrm{b}|/\tau_0=10^{-3}$. The onset time of the perturbation is denoted $t_\mathrm{pert}$. The evolution in pulse width is initially not smooth due to the spatial discretisation of the numerical solution, which allows only for approximate determination of the pulse tip and tail positions.}
  \label{fig:growthrate1}
\end{figure*}

The evolution of the pulse shape is best observed in the coordinate system that moves with the pulse tip at its reference speed $v_\mathrm{r,ss}$. Figure \ref{fig:pulseshape} shows contours of slip rate in the transformed coordinate system $\big((x-v_\mathrm{r,ss}t)/L^*, t/T^*\big)$ for the two simulations presented in Figure \ref{fig:pertexample1}. When the perturbation is negative (Figure \ref{fig:pulseshape}a), the pulse width reduction is initially driven by an acceleration of the trailing edge (healing front), and subsequently by a deceleration of the tip. The acceleration of the healing front initiates when the shear wave emitted from the pulse tip at the location of the perturbation reaches the trailing edge of the pulse. The overall pulse width reduces in a nonlinear manner over time, and the pulse arrests abruptly. When the perturbation is positive (Figure \ref{fig:pulseshape}b), an acceleration of the pulse tip is first observed, followed by a deceleration of the trailing edge. The pulse tip speed gradually approaches the shear wave speed. After a critical time of the order of $\sim20T^*$, the trailing edge accelerates again and further propagates at a speed greater than the

\newpage

\begin{figure*}[b]
  \centering
  \includegraphics{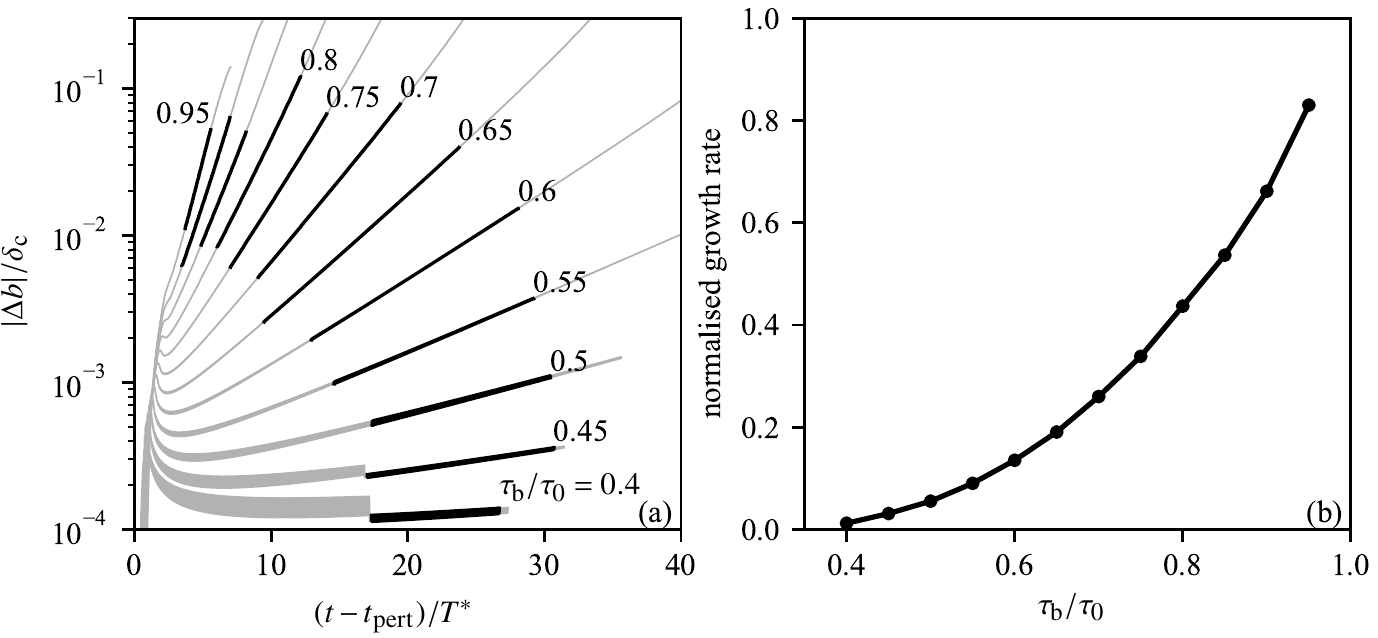}
  \caption{(a) Time evolution of slip perturbations following a negative background stress perturbation, for a range of reference background stresses $\tau_\mathrm{b}/\tau_0$. (b) Exponential growth rate of the slip perturbations as a function of the reference stress. The growth rate was computed using a least-square fit to a straight line of the data subset highlighted in black on the left panel.}
  \label{fig:slip_pert_taub}
\end{figure*}

\par\noindent initial $v_\mathrm{r,ss}$, but less than the tip speed. The slip pulse then becomes expanding.

\begin{figure*}
  \centering
  \includegraphics{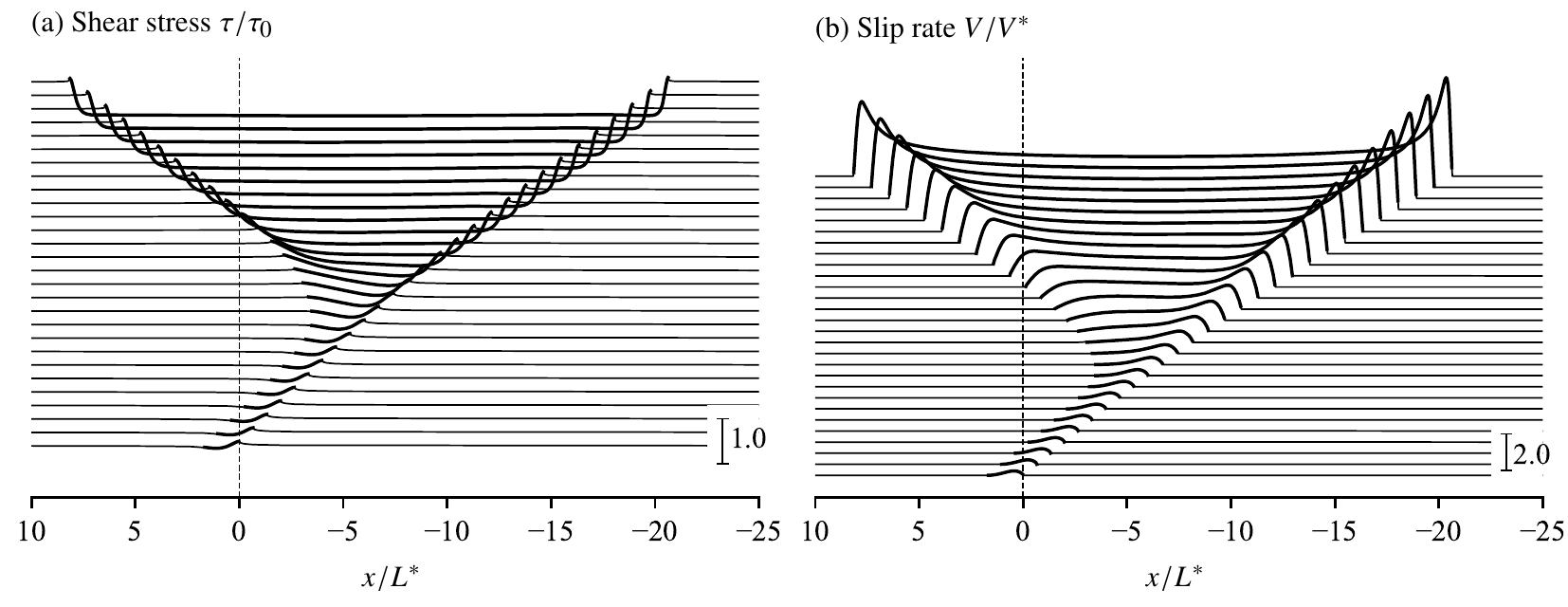}
  \caption{Snapshots of shear stress (a) and slip rate (b) profiles for a slip pulse propagating across a positive background stress perturbation. The initial steady state pulse is generated with a background stress $\tau_\mathrm{b}/\tau_0=0.9$, diffusivity ratio $\alpha_\mathrm{hy}/\alpha_\mathrm{th}=1$ and $h/h_\mathrm{dyna}=1$. The perturbation amplitude is $0.001\tau_0$, with a half-sinusoidal shape of width $L/3$ centered at $x=0$ (dotted line). In all the plots, thick lines mark the positions where slip rate is nonzero. Snapshots are shown at regular time intervals of $\approx 0.9T^*$.}
  \label{fig:pulse2crack}
\end{figure*}

The time evolution of perturbations in normalised pulse width, peak slip rate and slip is given in Figure \ref{fig:growthrate1}. Regardless of the sign of the perturbation in background stress, all the perturbed quantities appear to grow exponentially with time since the perturbation onset $t_\mathrm{pert}$ at which the pulse tip enters the perturbed region, until either the complete arrest of the pulse (Figure \ref{fig:growthrate1}a) or the transition to an expanding pulse (Figure \ref{fig:growthrate1}b, at $(t-t_\mathrm{pert})/T^*\gtrsim 25$). The amplitude of all the normalised variables is initially of the same order of magnitude as the perturbation in background stress, and all grow at approximately the same exponential rate. As shown in Figure \ref{fig:slip_pert_amplitude}, the amplitude of the perturbation impacts only the initial jump in the perturbed variables but does not modify the growth rate itself.

\begin{figure*}
  \centering
  \includegraphics{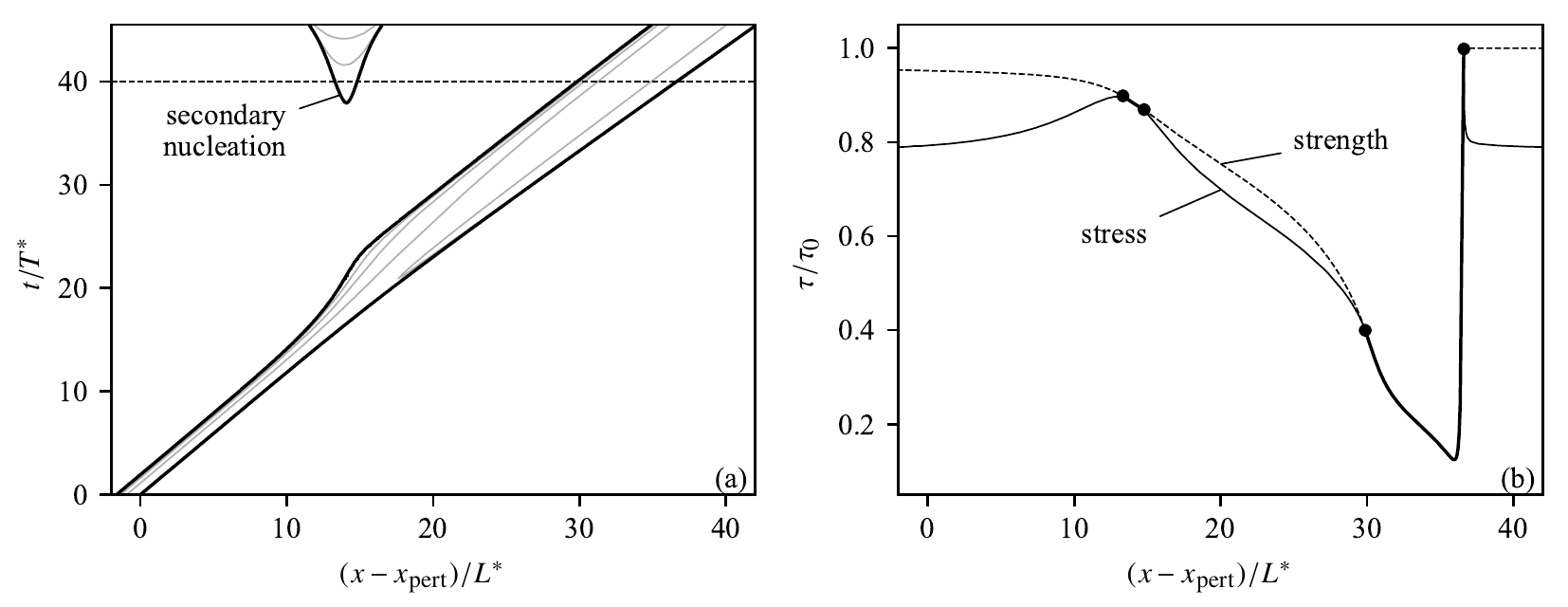}
  \caption{Effect of a positive background stress perturbation ($|\Delta\tau_\mathrm{b}|/\tau_0=10^{-3}$, starting at $x=x_\mathrm{pert}$) on a slip pulse propagating under an intermediate initial background stress $\tau_\mathrm{b}/\tau_0=0.783$. (a) Slip rate contours. Black line delineates the slipping patch (i.e., the $V=0$ contour), and grey lines are iso-$V$ contours logarithmically space between $V/V^*=0.01$ and $10$. (b) Shear stress (solid line) and strength (dotted line) profiles at $t/T^*=40$ (see dotted line in panel a). Thick black lines mark where slip rate is nonzero, and black dots mark the edges slipping patches.}
  \label{fig:renucleation}
\end{figure*}

The growth rate of slip perturbations following a negative stress perturbation is explored as a function of reference background stress in Figure \ref{fig:slip_pert_taub}. Reasonably accurate simulations can only be performed for $\tau_\mathrm{b}/\tau_0\gtrsim0.4$, because at lower stress levels the reference rupture speeds becomes too close to the shear wave speed (see Figure \ref{fig:pslip_taub}c). A clear trend of increasing growth rate with increasing reference shear stress is observed. This trend is not linear (Figure \ref{fig:slip_pert_taub}b): the growth rate approaches zero at low stress, and increases dramatically at high stress. In reference to Figure \ref{fig:pslip_taub}c, the overall trend implies that slow pulses with little net slip arrest more rapidly than faster ones.

At elevated background stress ($\tau_\mathrm{b}/\tau_0\gtrsim0.79$), the pulse response to positive perturbations is qualitatively different from that shown in Figures \ref{fig:pertexample1}c,d and \ref{fig:pulseshape}b. Figure \ref{fig:pulse2crack} shows a series of snapshots of shear stress and slip rate for a pulse propagating under a background stress $\tau_\mathrm{b}/\tau_0=0.9$ and perturbed at $x=0$ with a half-sine of width $(1/3)L/L^*$ and amplitude $\Delta\tau_\mathrm{b}/\tau_0=10^{-3}$. The tip of the pulse accelerates, and the tail decelerates and starts propagating in the negative $x$ direction, leaving an expanding region of nonzero slip rates across the crack line. The pulse-like rupture effectively transitions to a crack-like rupture.

At intermediate background stress ($\tau_\mathrm{b}/\tau_0 \sim 0.78$ for $\alpha_\mathrm{hy}/\alpha_\mathrm{th}=1$ and $h/h_\mathrm{dyna}=1$), a positive stress perturbation produces a complex rupture pattern, shown in Figure \ref{fig:renucleation}. The slip pulse initially transitions to a self-similar expanding pulse. At later times, a new, crack-like, rupture appears near the location where this transition occurred (Figure \ref{fig:renucleation}a). A plot of the shear stress and strength profiles (Figure \ref{fig:renucleation}b) reveals that the secondary nucleation is driven by the combination of a reduced strength in the wake of the pulse (although offset by strength recovery driven by pore fluid diffusion), and an increased backstress due to the expansion of the pulse. The net slip due to the expanding pulse increases approximately linearly with increasing propagation distance, so that the shear stress around the transition point from steady-state to expanding pulse is expected to increase logarithmically with time, and secondary nucleation ensues.

The process by which secondary nucleation might proceed is illustrated in detail in Figure \ref{fig:stressbuildup}, which shows (a) the stress profiles, and (b) the maximum stress perturbation behind the pulse (as well as the net slip perturbation) as a function of time for a simulation with positive stress perturbation and an initial background stress $\tau_\mathrm{b}/\tau_0=0.77$. Although secondary nucleation was not observed within the time frame of that simulation, a clear progressive increase in stress is observed around the location of the transition from steady-state to expanding pulse (see stress profiles inside the box in Figure \ref{fig:stressbuildup}(a)). A similar mechanism for the secondary rupture nucleation in the wake of expanding primary pulse, pulse or crack depending on the background stress level, have been described by \citet{gabriel12} for a fault with a velocity-weakening friction. This increase slows with increasing time and propagation distance (Figure \ref{fig:stressbuildup}(b)), but does not stabilise. This logarithmic increase in stress is expected if the net slip behind the pulse increases linearly with propagation distance, which seems to be the case here.

In summary, the numerical results presented above indicate that pulse-like ruptures driven by thermal pressurisation of pore fluids are unstable to infinitesimal perturbations. The growth of slip rate, slip and pulse width perturbations is initially exponential, and of the same sign as the initial stress perturbation. When that perturbation is negative, the slip pulse eventually stops, and does so abruptly. When the perturbation is positive, depending on the initial uniform background stress, the slip pulse grows and transitions to a self similar expanding pulse (at low stress) or an expanding crack-like rupture (at high stress). Because expanding pulses lead to an increasing net slip with increasing propagation distance, secondary nucleation is observed at the location of the perturbation at intermediate background stress.

\section{General stability criterion}
\label{sec:EoM}

The numerical results clearly indicate that steady-state slip pulses driven by thermal pressurisation are unstable. How general is this result? In this Section, an approximate equation of motion for dynamic pulse-like ruptures is established, and utilised to determine a general stability criterion for steady-state slip pulses.

\subsection{Slip pulse elastodynamics}

In the elastodynamic equilibrium equation
\begin{linenomath}
  \begin{equation}
    \tau(x,t) = \tau_\mathrm{b} -\frac{\mu}{2c_\mathrm{s}}V(x,t) + \phi(x,t), \label{eq:el}
  \end{equation}
\end{linenomath}
the stress redistribution functional has a form \citep{cochard97}
\begin{linenomath}
  \begin{equation}
    \phi(x,t) = \frac{\mu}{2\pi}\frac{\partial}{\partial x}\int_{-\infty}^{t}\int_{-\infty}^{+\infty}M\left(\frac{x-x'}{c_\mathrm{s}(t-t')}\right)\frac{\partial\delta/\partial x'(x',t')}{t-t'}dx'dt'. \label{eq:phi}
  \end{equation}
\end{linenomath}
The kernel $M(u)$ assumes a simple form for anti-plane deformation:
\begin{linenomath}
  \begin{equation}
    M(u) = \mathcal{H}(1-u^2)\sqrt{1-u^2}, \label{eq:M3}
  \end{equation}
\end{linenomath}
where $\mathcal{H}$ is the Heaviside function.

Consider a rupture in a form of a slip pulse of length $L(t)$ and
total accumulated slip (dislocation) $b(t)$, the motion of which is specified
by the coordinate of its tip, $x=\xi(t)$, advancing at generally
non-uniform speed $v_\mathrm{r}(t)=\dot{\xi}(t)$. On spatial scales much
larger than $L$, the pulse is seen as a singular dislocation
\begin{linenomath}
  \begin{align}
    |x-\xi(t)|\gg L(t): & \quad\delta(x,t)=b(\theta(x))\,\mathcal{H}(\xi(t)-x)\label{eq:dislocation}\\
                        & \quad\frac{\partial\delta}{\partial x}(x,t)=-b(\theta(x))\,\delta_\mathrm{Dirac}(\xi(t)-x)\nonumber\\
                        & \qquad\qquad +\frac{db(\theta(x))}{dx}\,\mathcal{H}(\xi(t)-x)\label{eq:gradient}                          
  \end{align}
\end{linenomath}
where $\theta(x)$ is the arrival time of the pulse at the position
$x$ (i.e., $\xi(\theta(x))=x$). Substituting this into \eqref{eq:el}
and \eqref{eq:phi},  moving $\partial/\partial x$ under the integral
in \eqref{eq:phi} yields after some manipulations (Appendix \ref{ax:phi})
\begin{linenomath}
  \begin{equation} \tau(x,t)-\tau_\mathrm{b}=\phi(x,t)-\frac{\mu}{2c_\mathrm{s}}v_\mathrm{r}(t)\,b(\theta(x))\,\delta_\mathrm{Dirac}(\xi(t)-x)\label{eq:el'}
  \end{equation}
\end{linenomath}
where
\begin{linenomath}                       
  \begin{align}
    \phi(x,t) &= \phi_\mathrm{Dirac}(x,t)+\phi_{H}(x,t),\label{eq:phi''}\\
    \phi_\mathrm{Dirac}(x,t) &= -\frac{\mu}{2\pi c_\mathrm{s}}\int_{-\infty}^{t}\frac{dM\left(\bar{u}\right)}{d\bar{u}}\frac{b(t')\,dt'}{(t-t')^{2}},\label{eq:phiD}\\
    \phi_{H}(x,t) &= -\frac{\mu}{2\pi}\int_{-\infty}^{t}M\left(\bar{u}\right)\frac{db}{dt'}\frac{dt'}{x-\xi(t')},\label{eq:phiH}
  \end{align}
\end{linenomath}
with 
\begin{linenomath}
  \begin{equation}
    \bar{u}\equiv\frac{x-\xi(t')}{c_\mathrm{s}(t-t')}.\label{eq:ubar}
  \end{equation}
\end{linenomath}
Terms $\phi_\mathrm{Dirac}$ and $\phi_{H}$ in the stress transfer functional
$\phi=\phi_\mathrm{Dirac}+\phi_{H}$ correspond to the singular (Dirac) and
the non-singular (step function $H$) terms in the slip gradient (Equation
\ref{eq:gradient}), respectively. 

Since the region of applicability of ``dislocation approximation''
\eqref{eq:el'}, $|x-\xi(t)|\gg L(t)$, excludes $x=\xi(t)$, the singular
($\delta_\mathrm{Dirac}$) term in \eqref{eq:el'} is of no consequence, and will be dropped in the foregoing, i.e., 
\begin{linenomath}
  \begin{equation}
    |x-\xi(t)|\gg L(t):\quad\quad\tau(x,t)-\tau_\mathrm{b}=\phi(x,t).\label{eq:el''}
  \end{equation}
\end{linenomath}

\subsection{Intermediate field of pulse}

Let us introduce the coordinate frame moving with the dislocation
(or the advancing front of the pulse), 
\[
X=\xi(t)-x.
\]
In the case of a steady (``steady state'') pulse motion, $\dot{b}=\dot{v}_\mathrm{r}=0$,
one finds that $\bar{u}=v_\mathrm{r}/c_\mathrm{s}-X/c_\mathrm{s}(t-t')$, and, in the case
of anti-plane slip, \eqref{eq:M3}, one recovers from \eqref{eq:el'} and
\eqref{eq:phiD} \citeauthor{weertman80}'s {[}1980{]} solution for a
subsonic dislocation,
\begin{linenomath}
  \begin{equation}
    \tau(x,t)-\tau_\mathrm{b}=\phi_\mathrm{ss}(X;v_\mathrm{r},b)\equiv-\frac{\mu\sqrt{1-v_\mathrm{r}^{2}/c_\mathrm{s}^{2}}}{2\pi}\frac{b}{X}.\label{eq:ss}
  \end{equation}
\end{linenomath}

\begin{figure*}[t!]
  \centering
  \includegraphics{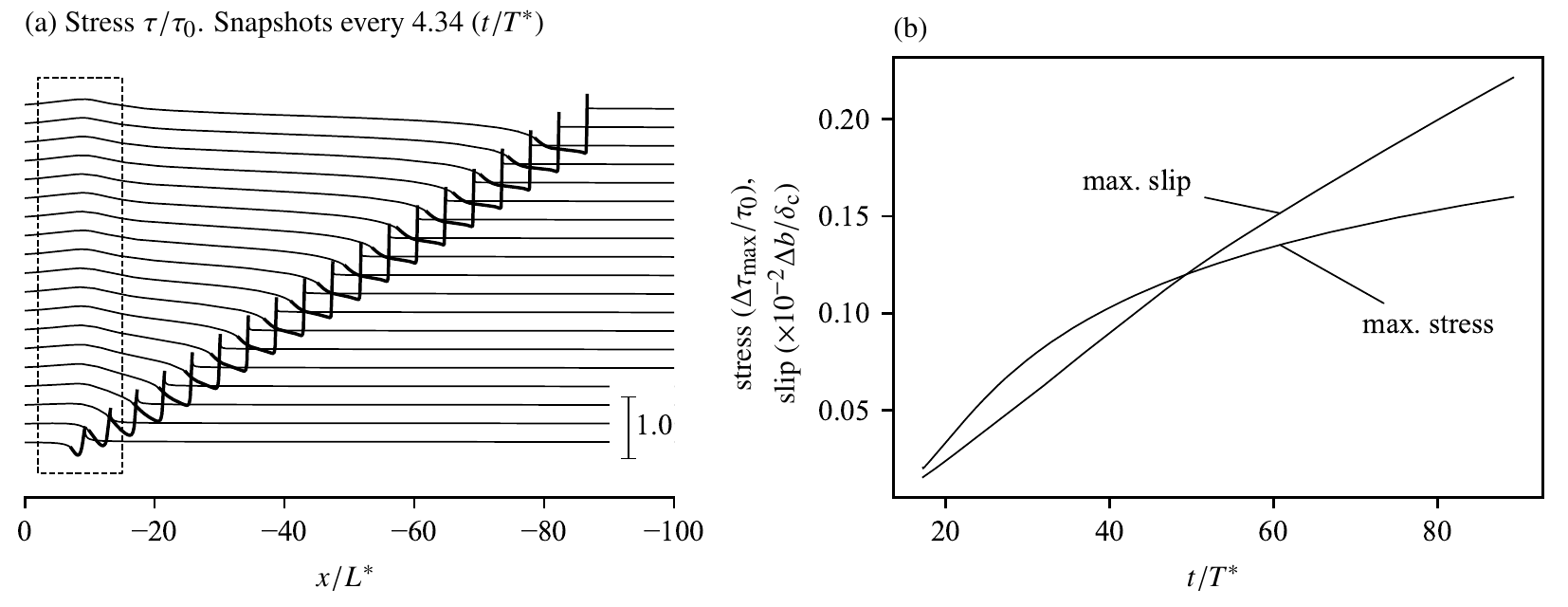}
  \caption{Stress buildup due to the transition to an expanding pulse. Simulation run using $\tau_\mathrm{b}/\tau_0=0.77$, and a positive stress perturbation located at $x=0$. (a) Snapshots of stress profiles. Thick lines highlight the slipping patch. Box highlights the stress buildup around the transition point. (b) Evolution of the maximum stress perturbation behind the pulse and maximum slip perturbation as a function of time.}
  \label{fig:stressbuildup}
\end{figure*}

In the general (non-steady) case, the Weertman's solution \eqref{eq:ss} with instantaneous values of $b(t)$ and $v_\mathrm{r}(t)$ gives the leading order term in the near field of a dislocation \citep[e.g.][]{eshelby53,markenscoff80,pellegrini10}. This near field can be defined by distances $X$ that are much smaller than a length scale $L_\mathrm{out}$ which characterises the unsteady motion of the pulse. For instance, if a dislocation accelerates or decelerates over a time scale $T_\mathrm{out}$ (e.g., $v_\mathrm{r}/\dot{v_\mathrm{r}}$ or $b/\dot{b}$), the associated length scale would be $L_\mathrm{out}=v_\mathrm{r}T_\mathrm{out}$. When considering a slip pulse, the approximation to a dislocation (Equation \ref{eq:el'}) only holds at distances much larger than the pulse length $L$, so that the near field of a dislocation corresponds to an intermediate field ($L\ll|X|\ll L_\mathrm{out}$) for a pulse, as long as the pulse ``inner'' lengthscale $L$ and the ``outer'' lengthscale $L_\mathrm{out}$ are separable, 
\begin{linenomath}
  \begin{equation}
    L\ll L_\mathrm{out}.\label{eq:bubble}
  \end{equation}
\end{linenomath}

Furthermore, as shown by \citet{eshelby53} on a particular example
of accelerating dislocation motion, and by \citet{markenscoff80,callias88,ni09}
in the case of general motion $\xi(t)$ of a dislocation of invariant
strength $b(t)=\text{constant}$, the next order term in the near field expansion
of a non-uniformly moving dislocation is logarithmically singular.
An extension of the results of \citet{ni09} for the $\phi_\mathrm{Dirac}$-expansion
to the general case with arbitrary time-dependencies $\xi(t)$ and
$b(t)$ yields (see Appendices \ref{ax:phiD} and \ref{ax:alternative} for details) 
\begin{linenomath}
  \begin{align}
    \phi_\mathrm{Dirac}(X,t)&\simeq\phi_\mathrm{ss}(X;v_\mathrm{r}(t),b(t))\nonumber\\
    &\quad+\frac{\mu}{4\pi c_\mathrm{s}b(t)}\,\frac{d}{dt}\left[b^{2}(t)\frac{v_\mathrm{r}(t)/c_\mathrm{s}}{\sqrt{1-v_\mathrm{r}^{2}(t)/c_\mathrm{s}^{2}}}\right]\ln\left|\frac{X}{L_\mathrm{out}}\right|. \label{eq:phiD''}
  \end{align}
\end{linenomath}
Note that we nondimensionalised $X$ under the logarithm with ``outer'' lengthscale $L_\mathrm{out}$, but could have used other similar length for this purpose. This is due to the fact that any such scaling length contributes only to the high order terms, $O(X^{0})$, in the expansion. Admittedly, it would be advantageous to include these higher order terms to improve the approximation provided by this expansion (especially in view of the equation of motion discussed in the forthcoming). However, the actual expression for the $O(X^{0})$ correction is very cumbersome, and appears to depend on the history of slip \citep{callias88,ni08}, and, consequently, is not included in \eqref{eq:phiD''}.

To find the near-field expansion for $\phi_{H}$ (Equation \ref{eq:phiH}), we first write
\begin{linenomath}
  \begin{equation}  \phi_{H}(x,t)=\int_{-\infty}^{t}(\partial\phi_{H}(x,t)/\partial t)dt,
  \end{equation}
\end{linenomath}
where 
\begin{linenomath}
  \begin{equation}
    \frac{\partial\phi_{H}(x,t)}{\partial t}=\frac{\mu}{2\pi c_\mathrm{s}}\int_{-\infty}^{t}\frac{dM\left(\bar{u}\right)}{d\bar{u}}\frac{db/dt'}{(t-t')^{2}}dt',\label{eq:phiH'}
  \end{equation}
\end{linenomath}
with $\bar{u}$ defined in \eqref{eq:ubar}. Exploiting the similarity between the integrals in expressions for $\phi_\mathrm{Dirac}$ (Equation \ref{eq:phiD}) and $\partial\phi_{H}/\partial t$ (Equation \ref{eq:phiH'}), and, in view of the $\phi_\mathrm{Dirac}$-expansion (equation \ref{eq:phiD''}), the leading term in the expansion for $\partial\phi_{H}/\partial t$ is given by 
\begin{linenomath}
  \begin{equation}
    \frac{\partial\phi_{H}(x,t)}{\partial t}\simeq-\phi_\mathrm{ss}\left(X;v_\mathrm{r}(t),\frac{db}{dt}\right).
  \end{equation}
\end{linenomath}
Integrating, we have 
\begin{linenomath}
  \begin{equation}
    \phi_{H}(x,t)\simeq\frac{\mu}{2\pi}\frac{\sqrt{1-v_\mathrm{r}^{2}(t)/c_\mathrm{s}^{2}}}{v_\mathrm{r}(t)}\,\frac{db}{dt}\,\ln\left|\frac{X}{L_\mathrm{out}}\right|,\label{eq:phiH''}
  \end{equation}
\end{linenomath}
where, once again, the choice of normalising lengthscale under the logarithm ($L_\mathrm{out}$) is, apart from the order of magnitude considerations, somewhat arbitrary. 

Combining \eqref{eq:phiD''} and  \eqref{eq:phiH''}, and simplifying, the near field expansion of the stress perturbation due to a moving dislocation takes the form
\begin{linenomath}
  \begin{align}
    \phi(x,t)&\simeq\phi_\mathrm{ss}(X;v_\mathrm{r}(t),b(t))\nonumber\\
    &+\frac{\mu}{2\pi}\frac{1}{v_\mathrm{r}(t)(1-v_\mathrm{r}^{2}(t)/c_\mathrm{s}^{2})^{1/4}}\frac{d}{dt}\left[\frac{b(t)}{(1-v_\mathrm{r}^{2}(t)/c_\mathrm{s}^{2})^{1/4}}\right]\ln\left|\frac{X}{L_\mathrm{out}}\right|.\label{eq:phi_int}
  \end{align}
\end{linenomath}

\subsection{Equation of motion of a moving dislocation}

In view of (\ref{eq:bubble}), the stress $\tau(x,t)$ at intermediate distances from the pulse is approximately given by that of the steady-state dislocation with instantaneous strength $b(t)$, moving, as dictated by the steady-state pulse solution, at $v_\mathrm{r}\simeq v_\mathrm{r,ss}(b(t))$ within the ``transient'' background stress field, $\tau_\mathrm{b,ss}(b(t))$, i.e.,
\begin{linenomath}
  \begin{equation}
    L\ll|X|\ll L_{out}:\quad\tau(x,t)-\tau_\mathrm{b,ss}(b(t))=\phi_\mathrm{ss}(X;v_\mathrm{r,ss}(b(t)),b(t)).\label{eq:int'}
  \end{equation}
\end{linenomath}
This type of approximation of the intermediate field of the unsteady dislocation appears to have been first suggested by \citet[p. 251]{eshelby53} when treating the particular example of a constant-strength dislocation accelerating from rest. Comparing \eqref{eq:int'} to \eqref{eq:el''} with \eqref{eq:phi_int}, leads to an ordinary differential equation (ODE) describing the evolution of total slip $b(t)$ accrued in an unsteadily propagating pulse:
\begin{linenomath}
  \begin{equation}
    \tau_\mathrm{b}-\tau_\mathrm{b,ss}(b(t))\simeq-\frac{\mu}{2\pi}\frac{1}{(1-v_\mathrm{r}^{2}/c_\mathrm{s}^{2})^{1/4}\,v_\mathrm{r}}\frac{d}{dt}\left[\frac{b}{(1-v_\mathrm{r}^{2}/c_\mathrm{s}^{2})^{1/4}}\right]\ln\left[\frac{L}{L_\mathrm{out}}\right],\label{eq:EofM}
  \end{equation}
\end{linenomath}
where $v_\mathrm{r}=v_\mathrm{r,ss}(b(t))$ and $L=L_\mathrm{ss}(b(t))$ are the steady-state pulse velocity and width, respectively. In arriving to the form \eqref{eq:EofM}, the slowly space-varying $\ln|X/L_\mathrm{out}|$ term in \eqref{eq:int'} was approximated by its value at distances of few $L$ away from the trailing edge of the pulse. Equation \eqref{eq:EofM} can be regarded as an ``equation of motion'' of an unsteady pulse, since once its solution $b=b(t)$ is known, the corresponding pulse trajectory follows by integration of $d\xi/dt=v_\mathrm{r,ss}(b(t))$.

In summary, the derived equation of motion relies on separation of spatial scales associated with the slip development within the pulse ($L$) and the evolution of the pulse net characteristics ($L_\mathrm{out}$), respectively, ($L\ll L_\mathrm{out}$). This scale separation allows to approximate unsteady pulse solution at a given instant of time by the steady-state solution (for a steadily propagating pulse) corresponding to the instantaneous value of total accrued slip $b(t)$, and other net pulse characteristics uniquely defined by the value of $b$ (i.e., $v_\mathrm{r}=v_\mathrm{r,ss}(b)$,
$L=L_\mathrm{ss}(b)$, etc.). The evolution of the pulse ``state variable'' $b(t)$ is specified by the equation of motion \eqref{eq:EofM}.

\subsection{Stability of steady-state pulse}

Equation \eqref{eq:EofM} allows to easily address the question of stability of a steady-state pulse solution (i.e., a solution of \eqref{eq:EofM} with $db/dt=0$). If the rupture velocity of a steady-state pulse monotonically increases with total slip, $dv_\mathrm{r,ss}/db\geq0$ and limited by $c_\mathrm{s}$ (see, for example, \citet{garagash12} for steady rupture pulses driven by thermal pressurisation of pore fluid, and our Figure \ref{fig:pslip_taub}b,c), and in view of $L_\mathrm{out}\gg L$, the right hand side of \eqref{eq:EofM} is a positive multiple of $db/dt$. It then follows from \eqref{eq:EofM} that the sign of $db/dt$ is set by that of $\tau_\mathrm{b}-\tau_\mathrm{b,ss}(b)$, and, thus, a steady-state solution with $b(t)=b_{0}$ is stable to small perturbations if and only if the steady-state value of the background stress increases with slip, $(d\tau_\mathrm{b,ss}/db)_{|b=b_{0}}>0$. (Interestingly, a similar stability condition was cited by \citet{rosakis01} without a proof). 

For faults that dynamically weaken with slip, smaller levels of background stress are not inconsistent with larger required slip (and more pronounced weakening that comes with it) to drive a pulse rupture. We, therefore, expect the condition
\begin{linenomath}
  \begin{equation}
    d\tau_\mathrm{b,ss}/db\leq0,\label{eq:inst}
  \end{equation}
\end{linenomath}
to be satisfied for a number of realistic constitutive laws (such as weakening by thermal pressurisation, as shown in Figure \ref{fig:slip_pert_taub}b), and thus inherently unstable steady-state pulse solutions. Indeed, initially steadily propagating slip pulses in a number of numerical studies utilising different models for the fault strength  \citep[e.g.,][]{perrin95,beeler96,zheng98,noda09,gabriel12}, eventually become unsteady, either growing (accelerating and accruing increasing levels of slip with distance travelled) or dying (shrinking and decelerating).

\subsection{Perturbation growth rate}

Let us write the equation of motion \eqref{eq:EofM} in a shorthanded form $\tau_\mathrm{b}-\tau_\mathrm{b,ss}(b)=\mu \Psi(b)\,db/dx$, where 
\begin{linenomath}
  \begin{equation} \label{eq:psib}
    \Psi(b)=\frac{1}{2\pi}\frac{1}{(1-v_\mathrm{r}^{2}/c_\mathrm{s}^{2})^{1/4}}\frac{d}{db}\left[\frac{b}{(1-v_\mathrm{r}^{2}/c_\mathrm{s}^{2})^{1/4}}\right]\ln\left[\frac{L_\mathrm{out}}{L}\right]
\end{equation}
\end{linenomath}
and, as before, $v_\mathrm{r}=v_\mathrm{r,ss}(b)$ and $L=L_\mathrm{ss}(b)$. Nondimensional
function $\Psi(b)$ is positive when, e.g., the steady-state rupture velocity is increasing with increasing net slip, as in the case of steady-state pulses driven by thermal pressurisation. Regardless of the sign of $\Psi(b)$, any small perturbation $\Delta b_\mathrm{ini}=(b-b_0)_\mathrm{ini}$ from the
steady-state pulse propagation with $b=b_0$, will initially evolve
with the propagated distance $x$ as
\begin{linenomath}
  \begin{equation}
    \Delta b=\Delta b_\mathrm{ini}\exp\left(-\frac{1}{\mu \Psi}\frac{d\tau_\mathrm{b,ss}}{db}x\right),\label{eq:pert}
  \end{equation}
\end{linenomath}
where $\Psi$ and $d\tau_\mathrm{b,ss}/db$ are evaluated at the baseline
state $b=b_0$. The exponential form \eqref{eq:pert} is qualitatively consistent with the numerical simulation using thermal pressurisation as a weakening mechanism (Figures \ref{fig:growthrate1} and \ref{fig:slip_pert_taub}a).

In the event when the slip perturbation is seeded by a background stress perturbation $\Delta\tau_\mathrm{b}$ localised in space over the dimension $\Delta x$, as is the case in our numerical simulations, the corresponding level of equivalent ``initial'' slip perturbation $\Delta b_\mathrm{ini}$ (that will persist and evolve according to \eqref{eq:pert} for $x>\Delta x$) can be estimated as 
\begin{linenomath}
  \begin{equation}
    \Delta b_\mathrm{ini}\approx\frac{\Delta\tau_\mathrm{b}}{\mu\,\Psi}\Delta x.
  \end{equation}
\end{linenomath}
The perturbation exponential growth rate, given by
\begin{linenomath}
  \begin{equation}
    s = -\frac{v_\mathrm{r}}{\mu \Psi}\frac{d\tau_\mathrm{b,ss}}{db}, \label{eq:growthrate}
  \end{equation}
\end{linenomath}
is therefore expected to be independent of the (small) perturbation amplitude. These general observations are consistent with the numerical simulations, which show a linear scaling between the perturbation in stress and the resulting slip perturbation, and the independence of the growth rate on the stress perturbation amplitude (Appendix \ref{ax:ampl}, Figure \ref{fig:slip_pert_amplitude}).

\begin{figure}
  \centering
  \includegraphics{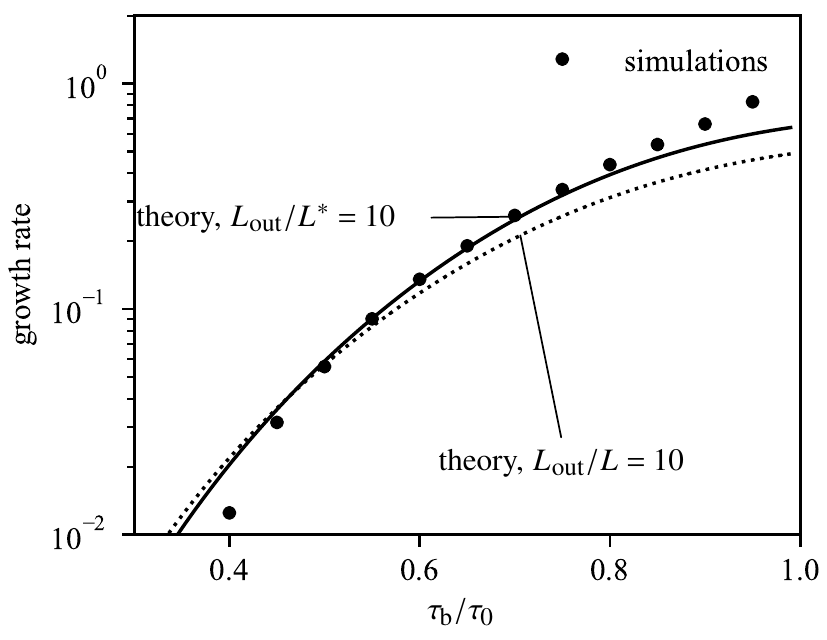}
  \caption{Comparison of perturbation growth rates from numerical simulations (dots) and theoretical estimates based on the slip pulse equation of motion (solid and dotted lines). The latter are computed using the relationships $\tau_\mathrm{b,ss}(b)$, $v_\mathrm{r,ss}(b)$ shown in Figure \ref{fig:pslip_taub} and Equation \eqref{eq:growthrate}, with either $L_\mathrm{out}/L^*=10$ and $L=L_\mathrm{ss}$ (solid line) or $L_\mathrm{out}/L=10$ (dotted line).}
  \label{fig:comparegrowth}
\end{figure}

The steady-state solutions presented in Section \ref{sec:TPnum} provide the relationships between $\tau_\mathrm{b,ss}$, $v_\mathrm{r,ss}$ and slip $b$  (see Figure \ref{fig:pslip_taub}) required to compute a theoretical estimate of the growth rate using Equation \eqref{eq:growthrate}, leaving only the ratio $L_\mathrm{out}/L$ as an unconstrained parameter. Using $L_\mathrm{out}/L=10$ or a constant $L_\mathrm{out}/L^*=10$ and the steady-state pulse width $L=L_\mathrm{ss}$ produces the results shown in Figure \ref{fig:comparegrowth} (dotted and solid lines, respectively), where the growth rates estimated from numerical simulations are also displayed for comparison. The agreement between theoretical and numerical estimates with either choice for $L_\mathrm{out}$ is very satisfactory, and illustrates the applicability of the pulse equation of motion \eqref{eq:EofM}. Since the ratio $L_\mathrm{out}/L$ only appears in the logarithmic term, the resulting growth rate is not very sensitive to the specific choice for this unconstrained quantity. It appears that choosing $L_\mathrm{out}$ to be several times larger than $L$ produces reasonable predictions, consistent with the assumption \eqref{eq:bubble}.

\subsection{Validity of equation of motion}

\begin{figure}
  \centering
  \includegraphics{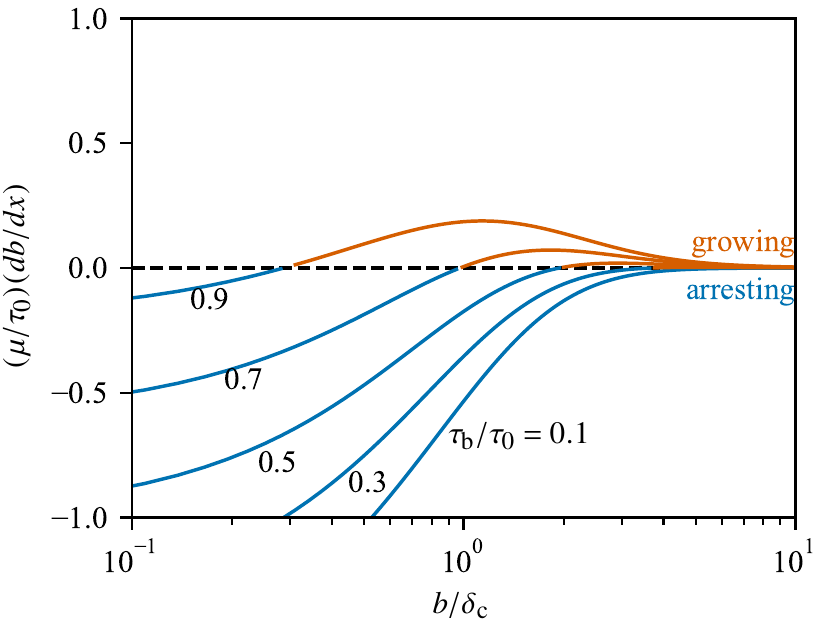}
  \caption{Scaled slip gradient as a function of slip (normalised by $\delta_\mathrm{c}$) for a range of background stresses, computed from the approximate equation of motion (Equation \ref{eq:EofM}), using thermal pressurisation-driven steady-state pulse characteristics with $h/h_\mathrm{dyna}=1$. In the computation of $\Psi(b)$ (Equation \ref{eq:psib}), a constant $L_\mathrm{out}/L=10$ was used.}
  \label{fig:eofmvalid}
\end{figure}

The key underlying assumption in our derivation of the approximate equation of motion for the slip pulse (Equation \ref{eq:EofM}) is that the pulse is in ``quasi-steady-state'', i.e., its characteristics (accrued slip $b$, length $L$ and speed $v_\mathrm{r}$) change slowly on the timescale of slip. This assumption can be translated in terms of propagation distance, since $b,L$ and $v_\mathrm{r}$ do not vary appreciably over propagation distances of the order of the pulse length $L$: the quasi-steady-state approximation is then valid as long as $|d(b/\delta_\mathrm{c})/d(x/L^*)|\ll1$, that is, $(\mu/\tau_0)|db/dx|\ll1$. This assumption can be validated from the equation of motion itself: indeed, using the notation introduced in Equation \eqref{eq:psib},
\begin{linenomath}
  \begin{equation}
    \frac{db}{dx} = \frac{\tau_\mathrm{b} - \tau_\mathrm{b,ss}(b)}{\mu\Psi(b)},
  \end{equation}
\end{linenomath}
which is a function of slip $b$ and stress $\tau_\mathrm{b}$, plotted in Figure \ref{fig:eofmvalid}. For elevated background stresses, around $\tau_\mathrm{b}/\tau_0=0.9$, the normalised slip gradient $(\mu/\tau_0)(db/dx)$ remains significantly less than unity. This is also the case throughout the regime of growing pulses (i.e., when $\tau_\mathrm{b}>\tau_\mathrm{b,ss}$). We therefore expect the equation of motion to provide an adequate description of the pulse dynamics under those conditions. For arresting pulses, the assumption of quasi-steady-state becomes invalid as slip decreases, with the magnitude of the slip gradient rapidly becoming of the order of unity, notably under low background stresses. This can be understood by considering that steady-state pulses associated with small slip correspond to elevated background stresses and low rupture speeds (Figure \ref{fig:pslip_taub}b,c): the regime of arresting pulses under very low stresses $\tau_\mathrm{b}\ll\tau_\mathrm{b,ss}$ is therefore too far from steady-state and the pulse is expected to arrest quickly compared to the duration of slip.

\section{Discussion and implications}
\label{sec:discussion}



The pulse equation of motion and stability analysis demonstrate that steady-state pulses are unstable if, e.g., $d\tau_\mathrm{b,ss}/db<0$ and $dv_\mathrm{r}/db>0$, or more generally if the exponential growth rate is positive, which, in view of \eqref{eq:growthrate} and \eqref{eq:psib}, corresponds to
\begin{linenomath}
  \begin{equation}
    \frac{d\tau_\mathrm{b,ss}}{d[b/(1-v_\mathrm{r}^2/c_\mathrm{s}^2)^{1/4}]}<0.
  \end{equation}
\end{linenomath}
This condition is satisfied for pulses driven by thermal pressurisation, and the numerical simulations confirm qualitatively and quantitatively this instability. In the light of these results, two key questions arise: what do steady-state pulse solutions tell us about the dynamics of rupture in general? What does the existence of unstable slip pulses imply for earthquake dynamics and strength of faults?


\subsection{Significance of steady-state pulse solutions}

Steady-state slip pulses arise spontaneously in fully dynamic rupture simulations when the nucleation and background stress conditions are at the transition between arresting and growing ruptures \citep{noda09,schmitt11,gabriel12}. Therefore, the conditions leading to the existence of steady-state solutions coincide with those allowing for the existence of sustained ruptures. In other words, steady-state pulse solutions inform us about the overall ``strength'' of an interface, in the sense that they provide us with the critical conditions required for ruptures to propagate beyond their nucleation patch.

Our results complement the framework provided by \citet{zheng98} who determined the critical background stress level ($\tau_\mathrm{pulse}$) separating the regime of exclusively pulse-like ruptures under low stress conditions and the regime where both crack and pulse rupture modes are possible under high stress conditions. Here we show both theoretically and numerically (for the case of thermal pressurisation) that the pulse mode of rupture exists within the entire range of background stress (low and high), while the dynamics of the pulse (spontaneous decay leading to arrest or spontaneous growth leading to either transition into crack-like rupture or nucleation of a secondary rupture in the pulse wake) can be extracted from a steady-state pulse analysis, like that conducted by \citet{garagash12} and summarised in Section \ref{sec:ss} for the case of thermal pressurisation. Although we do not establish conditions for prevalence of the pulse-like mode for ruptures driven by thermal pressurisation (such as the $\tau_\mathrm{pulse}$ threshold of \citet{zheng98} for velocity-weakening friction case), we suspect that the $\tau_\mathrm{pulse}$ threshold in this case would correspond to the minimum level of background stress at which the secondary rupture nucleated in the wake of growing primary pulse is crack-like. Therefore, the solution to the steady-state pulse problem associated with a particular constitutive behaviour provides a tool to determine the exact conditions (notably in terms of background stress) leading to the existence of sustained ruptures. Our analysis on the role of unstable slip pulses in controlling the growth of large scale ruptures is consistent with recent theoretical results from \citet{brener18}, who analysed numerically the stability of slip pulses driven by a nonlinear rate-dependent friction law. Numerical simulations indicate that such slip pulses are also unstable to small perturbations, and \citet{brener18} argue that such instabilities can be viewed as the nucleation process of large ruptures.

In the case of thermal pressurisation, the minimum dynamic strength is zero and thus there is no lower stress limit for the existence of dynamic steady-state slip pulses. Therefore, faults governed by thermal pressurisation have theoretically no ``strength'': thermal pressurisation allows for large enough pulses to propagate regardless of the initial background stress. However, theoretical slip pulses propagating under very low stress conditions bear large slip and slip rate, and require nucleation conditions characterised by either very high local stresses or large nucleation region (with modestly elevated stress). The question of the minimum stress required for ruptures to grow is therefore linked to the nucleation conditions of those ruptures. This was illustrated by \citet{gabriel12} in the context of a slip rate dependent constitutive law, who showed that the threshold background stress between arresting and growing pulses (i.e., the steady-state pulse regime) scales with the size of the nucleation patch used in their simulations. The nucleation conditions probably enforce the selection of a specific characteristic pulse width, stress drop and slip rate, and the background stress level outside the nucleation patch selects whether the rupture will become crack-like or an expanding pulse or decaying pulse, the boundary between the latter two regimes being determined by the steady-state stress for that pulse.


\subsection{Complexity of earthquake ruptures}

Our results show that unstable slip pulses can produce remarkably complex rupture events, even when the background stress and conditions are uniform (except for an infinitesimal perturbation). Rupture complexity is thus not systematically linked to complexity or heterogeneity in stress or strength conditions, but arise spontaneously when ruptures propagate as slip pulses.

\begin{figure}
  \centering
  \includegraphics{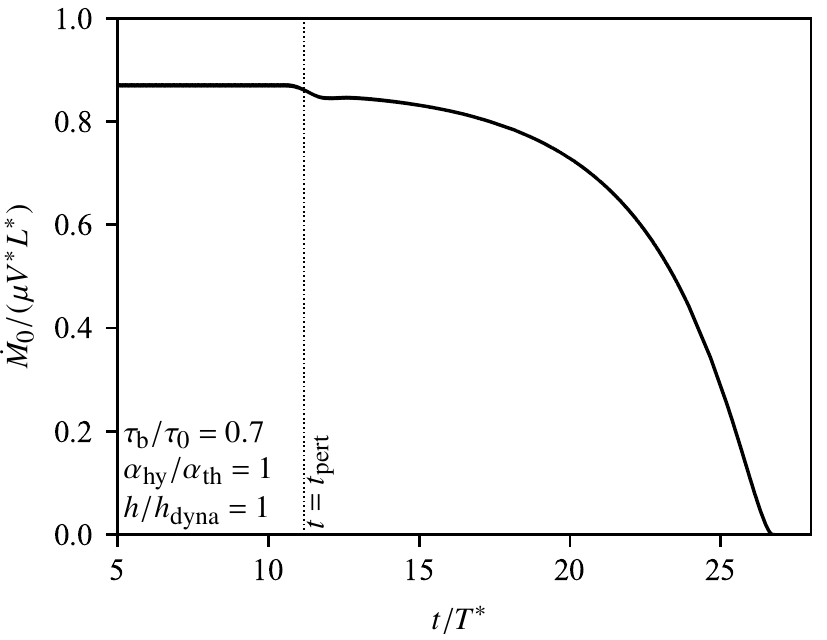}
  \caption{Normalised moment rate as a function of time for a slip pulse arresting due to a small negative stress perturbation centered at $t=t_\mathrm{pert}$ ($\Delta\tau_\mathrm{b}/\tau_0=-10^{-3}$).}
  \label{fig:momentrate}
\end{figure}

A notable feature observed for pulses arresting due to negative stress perturbations is that the arrest is abrupt. Negative perturbations in slip, slip rate and pulse width grow exponentially over time until the rupture stops. This is best illustrated by computing the (one dimensional) moment rate,
\begin{linenomath}
  \begin{equation} \label{eq:mdot}
    \dot{M}_0(t) = \mu\int_0^LV(X,t)X,
  \end{equation}
\end{linenomath}
which is shown in Figure \ref{fig:momentrate} for a pulse propagating at $\tau_\mathrm{b}/\tau_0=0.7$ and arresting due to a small negative perturbation. The moment rate is initially constant, corresponding to the steadily propagating pulse solution. Upon encountering the stress perturbation, moment rate decreases exponentially and drops abruptly to zero. Such rapid variations in moment rate are responsible for the radiation of high frequency waves in the far field, and we confirm here that such high frequencies associated with sudden rupture arrest can arise without strong stress or strength heterogeneities on the fault plane. Similar conclusios were established by \citet{cochard94} and \citet{gabriel12} in simulations using strong velocity weakening friction.

Another key feature associated with the slip pulse instability is the transition from pulse-like to crack-like rupture due to positive stress perturbations under high uniform background stress (Figures \ref{fig:pulse2crack} and \ref{fig:renucleation}). Here again, we observe a remarkable complexity that emerges spontaneously in the absence of any preexisting fault heterogeneities. Above a critical background stress (here $\tau_\mathrm{b}/\tau_0\gtrsim0.79$), the slip pulse transitions directly to an expanding crack. Under low stress conditions, numerical results indicate that growing ruptures become expanding pulses. One important consequence of the transition to expanding pulse is that as the pulse further propagates, the accrued slip grows approximately linearly with propagation distance and therefore we expect a logarithmic stress build-up near the starting point of growth. This is observed in our simulations when the background shear stress is close to the threshold for the direct transition into crack-like rupture (see Figures \ref{fig:renucleation}, \ref{fig:stressbuildup}).

What this transition illustrate is that unstable steady-state pulses evolve towards the most stable rupture mode, either self-similar pulse or expanding crack, according to the current background stress level. However, a peculiarity exhibited by our results is that rupture arrest is also a strong attractor (when perturbations are negative), so that a nascent slip pulse propagating in an overall high stress regime might arrest on its own if negative stress perturbations are encountered.








\section{Conclusions}

We performed numerical simulations and a theoretical analysis that demonstrate that steady-state slip pulses are unstable if the accrued slip (``dislocation'') is a decreasing function of the background stress, i.e., $d\tau_\mathrm{b}/db\leq0$. This instability condition is satisfied for slip pulses driven by thermal pressurisation of pore fluids. During instability, slip, slip rate and pulse width perturbations grow exponentially. If the initial stress perturbation leading to instability is negative, ruptures eventually arrest in an abrupt manner; conversely, if the stress perturbation is positive, rupture mode changes and transitions to a growing pulse (at low stress) or an expanding crack (at high stress). The growth rate of perturbations is predicted quantitatively by an approximate equation of motion for a dislocation with variable net slip (Equation \ref{eq:EofM}).

The regime of steady-state pulse solutions appears naturally in dynamic rupture simulations at the transition between spontaneously expanding ruptures (growing pulses) and spontaneously arresting ruptures, at a stress level that depends on the nucleation conditions. Once nucleation conditions are established, the steady-state pulse solution provides a stress limit below which ruptures will spontaneously stop, which is best considered as the ``strength'' of the interface \citep[e.g.,][]{lapusta03,rubinstein04,noda09}.

The unstable character of steady-state slip pulses generates a remarkable complexity of ruptures, including abrupt arrest, pulse to crack transitions and secondary rupture nucleation in the wake of a propagating pulse, even though stress and material parameters are homogeneous (nonwithstanding an infinitesimal perturbation) along the fault.  Pulse-like ruptures seem to be the main rupture mode for many crustal faults \citep{heaton90}, and it is therefore expected that earthquake dynamics along these faults is driven at least in part by spontaneous instabilities. One key consequence is that abrupt arrest of ruptures may not be the signature of strong preexisting stress of strength heterogeneities along faults. At this stage, it remains to be explored how slip pulse instabilities evolve along heterogeneous faults, and further work in this direction is currently conducted. Preliminary simulations suggest that transient pulses (i.e., non steady-state) continuously grow or shrink as they cross regions of high and low stress, respectively, their eventual arrest being dictated by finite amplitude stress perturbations.

\begin{acknowledgments}
  All authors contributed equally to the work, and are listed in alphabetical order. This work was supported by the UK Natural Environment Research Council through grant NE/K009656/1 to NB, by the Canada Natural Science and Engineering Research Council though Discovery grant 05743 to DIG and by MEXT KAKENHI Grant Number 26109007 to HN. The results of this paper can be reproduced by direct implementation of the analytical formulae and numerical methods described in the main text and appendices.
\end{acknowledgments}

\appendix

\section{Convolution kernels for fault strength and temperature}
\label{ax:convolution}

The convolution kernel $\mathcal{K}$ is given by \citep{rice06,garagash12}
\begin{linenomath}
  \begin{equation} \label{eq:K}
    \mathcal{K}(z;\chi) = \frac{\chi\mathcal{A}(z/(1+1/\sqrt{\chi})^2) - \mathcal{A}(z/(1+\sqrt{\chi})^2)}{\chi -1},
  \end{equation}
\end{linenomath}
if $\chi\ne1$, or by the limit of that expression as $\chi\rightarrow 1$ if $\chi=1$. The function $\mathcal{A}$ depends on the spatial distribution of strain rate across the fault, and for our choice of a Gaussian distribution we have
\begin{linenomath}
  \begin{equation} \label{eq:A}
    \mathcal{A}(z) = \frac{1}{\sqrt{\pi z +1}}.
  \end{equation}
\end{linenomath}

The pore pressure and temperature evolution on the fault plane ($y=0$) can be computed from the strength evolution as
\begin{linenomath}
  \begin{equation} \label{eq:p}
    p(0,t) = p_0 + (\sigma'_0 - \tau_\mathrm{f}/f),
  \end{equation}
\end{linenomath}
and
\begin{linenomath}
  \begin{equation} \label{eq:T}
    \Theta(0,t) = \Theta_0 + \frac{1}{\rho c h}\int_0^t\tau(t')V(t')\mathcal{A}\left(\frac{t-t'}{h^2/(4\alpha_\mathrm{th})}\right)dt'.
  \end{equation}
\end{linenomath}

\section{Numerical methods}

\subsection{Steady-state problem}
\label{ax:GC}

The method of solution for the steady-state pulse is the same as that employed by \citet{platt15b} in a more complex case (including thermal decomposition in addition to thermal pressurisation), and reviewed by \citet{viesca18}. For completeness, we provide here a description of the technique in the simple case of a finite pulse driven by thermal pressurisation only.


Normalising the slip by $\delta_\mathrm{c}$, time by $T^*$, stresses by $\tau_0$, distances by $\bar\mu\delta_\mathrm{c}/\tau_0$ and slip rate by $\delta_\mathrm{c}/T$, we rewrite the governing Equations \eqref{eq:pulse_eq} and \eqref{eq:tauf_integral} as
\begin{linenomath}
  \begin{equation} \label{eq:taussnorm}
    \tilde{\tau}(\tilde{x}) = \tilde{\tau}_\mathrm{b} + \frac{1}{2\pi \tilde{L}}\int_0^{\tilde{L}}\tilde{V}(\xi)\frac{d\xi}{\xi-\tilde{x}}
  \end{equation}
\end{linenomath}
and
\begin{linenomath}
  \begin{equation}\label{eq:constnorm}
    \tilde{\tau}_\mathrm{f}(\tilde{x}) = 1 - \int_0^{\tilde{L}}\mathcal{H}(\tilde{x}-\xi)\tilde{\tau}_\mathrm{f}(\xi)\tilde{V}(\xi)\mathcal{K}\left((\tilde{x}-\xi)\tilde{T}/\tilde{L};\chi\right)\frac{d\xi}{\tilde{L}},
  \end{equation}
\end{linenomath}
where normalised variables are denoted by a tilde, $\mathcal{H}$ is the Heaviside function, and $\chi=\alpha_\mathrm{hy}/\alpha_\mathrm{th}$. In Equation \eqref{eq:constnorm}, we changed the integration variable from time to space by noting that $\tilde{t}=\tilde{x}\tilde{T}/\tilde{L}$. The integrals in \eqref{eq:taussnorm} and \eqref{eq:constnorm} are further normalised using the transformed space coordinate $y=2\tilde{x}/\tilde{L}-1$, which results in
\begin{linenomath}
  \begin{align}
    \tilde{\tau}(y) &= \tilde{\tau}_\mathrm{b} + \frac{1}{\pi\tilde{L}}\int_{-1}^{1}\tilde{V}(y')\frac{dy'}{y'-y}, \\
    \tilde{\tau}_\mathrm{f}(y) &= 1 - \int_{-1}^1\mathcal{H}(y-y')\tilde{\tau}_\mathrm{f}(y')\tilde{V}(y')\mathcal{K}\big((y-y')T/2;\chi\big)dy'.
  \end{align}
\end{linenomath}
The condition \eqref{eq:kL} is similarly rewritten as
\begin{linenomath}
  \begin{equation}
    \int_{-1}^1\sqrt{\frac{1+y}{1-y}}\frac{d\tilde{\tau}}{dy}dy = 0.
  \end{equation}
\end{linenomath}

The idea now is to approximate the above integrals with Gauss-Chebyshev quadratures \citep{viesca18}. Because we expect the slip rate $\tilde{V}(y)$ to behave as $\sqrt{1\pm y}$ near $y\mp 1$ (i.e., square-root behaviour of the slip profile near the rupture tip and tail), we introduce the function $v(y)$ as
\begin{linenomath}
  \begin{equation}
    \tilde{V}(y) = v(y)\sqrt{1-y^2},
  \end{equation}
\end{linenomath}
which becomes the unknown (regular) function we are looking to approximate. Using the approximations
\begin{linenomath}
  \begin{equation}
    \int_{-1}^1\sqrt{1-y^2}f(Y-y)dy \approx \sum_{j=1}^nw_jf(Y_i-y_j),
  \end{equation}
\end{linenomath}
with
\begin{linenomath}
  \begin{equation*}
    \left\{\begin{array}{ll}
                                                                                           y_j&=\displaystyle \cos\left(\frac{\pi j}{n+1}\right),\\
                                                                                           Y_i&=\displaystyle \cos\left(\frac{\pi}{2}\frac{2j-1}{n+1}\right),\\
                                                                                           w_j&=\displaystyle (1-y_j^2)\frac{\pi}{n+1},\end{array}\right.
  \end{equation*}
\end{linenomath}
for $i=1,\ldots,n$, $j=1,\ldots,n+1$, and
\begin{linenomath}
  \begin{equation}
    \int_{-1}^1\sqrt{\frac{1+y}{1-y}}f(y)dy \approx \sum_{p=1}^nw_pf(y_p),
  \end{equation}
\end{linenomath}
with
\begin{linenomath}
  \begin{equation*}
    \left\{\begin{array}{ll}
                                                                                                      y_p&=\displaystyle \cos\left(\frac{\pi (2j-1)}{2n+1}\right),\\
                                                                                                      w_p&=\displaystyle \frac{2\pi(1+y_p)}{2n+1},\end{array}\right.
  \end{equation*}
\end{linenomath}
for $p=1,\ldots,n$, the governing equations become a linear system:
\begin{linenomath}
  \begin{align}
    \tilde{\tau}_i &= \tilde{\tau}_\mathrm{b} + \frac{1}{\pi \tilde{L}}\sum_{j=1}^n\frac{w_j}{\pi(y_j-Y_i)}v_j,\quad i=1,\ldots,n+1,\\
    \tilde{\tau}_i &= 1-\sum_{j=1}^n\mathcal{H}(Y_i-y_j)\tilde{\tau}_jv_j\mathcal{K}(\tilde{T}(Y_i-y_j)/2;\chi)w_j,\quad i=1,\ldots,n+1,\\
    0&=\sum_{p=1}^n w_p\frac{d\tilde{\tau}}{dy}\Big|_p,
  \end{align}
\end{linenomath}
where $\tilde{\tau}_i=\tilde{\tau}(Y_i)$, $\tilde{\tau}_j=\tilde{\tau}(y_j)$ and $v_j=v(y_j)$. In the system above, the normalised stress $\tilde{\tau}$ needs to be differentiated with respect to $y$, and evaluated at both sets of points $y_j$ and $Y_i$. Given the knowledge of the set of $\tilde{\tau_i}$, we use barycentric interpolation and Chebyshev differentiation matrices to compute
\begin{linenomath}
  \begin{align}
    \tilde{\tau}_j &= L_{ji}\tilde{\tau}_i,\\
    \frac{d\tilde{\tau}}{dy}\Big|_p &= D_{pj}L_{ji}\tilde{\tau}_i,
  \end{align}
\end{linenomath}
where $L_{ji}$ is an interpolation matrix \citep{viesca18} and $D_{pj}$ is a Chebyshev differentiation matrix \citep{trefethen00}, and we sum over repeated indices. In summary, we arrive at the following linear system:
\begin{linenomath}
  \begin{align}
    \tilde{\tau}_i &= \tilde{\tau}_\mathrm{b} - K_{ij}v_j/\tilde{L}, \label{eq:taui1}\\
    \tilde{\tau}_i &= 1-S_{ij}(L_{jk}\tilde{\tau}_kv_j), \label{eq:taui2}\\
    0&=w_pD_{pj}L_{pi}\tilde{\tau}_i, \label{eq:cond}
  \end{align}
\end{linenomath}
where $K_{ij}=w_j/(\pi(Y_i-y_j))$ and $S_{ij} = w_j\mathcal{H}(Y_i-y_j)\mathcal{K}((Y_i-y_j)T/2;\chi)$. Equating \eqref{eq:taui1} and \eqref{eq:taui2}, we obtain a total of $n+2$ equations, with $n+2$ unknowns that are $v_j$ ($j=1,\ldots,n$), $\tilde{L}$ and $\tilde{T}$. This system is solved using the Newton-Raphson iterative algorithm.

\subsection{Expression of stress transfer functional}
\label{ax:func}

Consider a spatial domain of length $\lambda$. Let $D_p(t)$ and $\dot{D}_p(t)$ denote the spatial discrete Fourier transform coefficients of the slip and slip rate perturbations, respectively, where indices $p$ correspond to wavenumbers $k_p=2\pi p/\lambda$. The discrete Fourier transform coefficients of the stress transfer functional $\phi$ are given by \citep{perrin95}
\begin{linenomath}
  \begin{equation} \label{eq:Fp}
    F_p(t) = -\frac{\mu|k_p|}{2}D_p(t) + \frac{\mu|k_p|}{2}\int_0^t W(|k_p|c_\mathrm{s}t')\dot{D}_p(t-t')dt',
  \end{equation}
\end{linenomath}
where $W(u)=\int_0^\infty \left(J_1(x)/x\right) dx$, and $J_1$ is the Bessel function of the first kind of order one. An inverse Fourier transform of $F_p(t)$ provides the value of $\phi$ in the space-time domain.

\subsection{Dynamic problem}
\label{ax:numerics}

The technique employed to solve the elastodynamic problem is essentially following the spectral boundary integral method of \citet{lapusta00}, adapted to our specific choice of constitutive behaviour (thermal pressurisation with constant friction coefficient). In this method, the dynamic stress transfer functional is evaluated in the Fourier domain, taking advantage of the efficiency of Fast Fourier Transform (FFT) algorithm.

The space domain is discretised into nodes $x_i=ih$, $i=1,\ldots,N$. Time is discretised into steps $t_n$, $n=0,\ldots,N_t$, with a constant spacing $\Delta t$. We denote with subscripts $i$ and superscripts $n$ the discretised variables at node $(x_i,t_n)$.

We first determine a steady-state solution for a uniform background stress and a given diffusivity ratio. The stress and slip rate distributions, $\tau_\mathrm{ss}$, $V_\mathrm{ss}$, are interpolated onto our regular grid at each node $(x_i,t_n)$, so that $\tau_{\mathrm{ss},i}(t_n)$ and $V_{\mathrm{ss},i}(t_n)$ are precomputed and stored a priori. At time $t_0$, we initialise the perturbations in slip ($\Delta\delta_i$), slip rate ($\Delta V_i$), stress ($\Delta\tau_i$) and strength ($\Delta\tau_{\mathrm{f},i}$) with zeros at all nodes.

Let us consider that all variables are known at a given time step $t_n$, including the entire slip rate perturbation history (and its Fourier coefficients, for use in the spectral boundary integral algorithm). The computation of variables at time step $t_{n+1} = t_n+\Delta t$ is conducted as follows:
\paragraph{1} Make a first estimate of the slip perturbation assuming a slip rate perturbation equal to that at time step $t_n$:
\begin{linenomath}
  \begin{equation}
    \Delta\delta^*_i = \Delta\delta_i^n + \Delta V_i^n \Delta t.
  \end{equation}
\end{linenomath}
\paragraph{2} Estimate the perturbation in dissipation rate (denoted $\Delta(\tau V)$) for the interval $[t_n, t_{n+1}]$ as
\begin{linenomath}
  \begin{align}
    \Delta(\tau V)^{n+1/2}_i &= \Delta V_i^n \Delta\tau_i^n + \frac{1}{2}(\Delta V_{\mathrm{ss},i}^n + \Delta V_{\mathrm{ss},i}^{n+1})\Delta\tau_i^n \nonumber\\
    &\qquad \qquad + \frac{1}{2}(\Delta \tau_{\mathrm{ss},i}^n + \Delta \tau_{\mathrm{ss},i}^{n+1})\Delta V_i^n,
  \end{align}
\end{linenomath}
and compute the perturbation in strength as
\begin{linenomath}
  \begin{equation}\label{eq:Dtauf1}
    \Delta\tau_{\mathrm{f},i}^* = \sum_{k=1}^n\Delta(\tau V)_i^{k+1/2} \mathcal{K}(t_n-t_k+\Delta t/2;\chi)\Delta t,
  \end{equation}
\end{linenomath}
which corresponds to a mid-point approximation of the integral in \eqref{eq:dtauf}. The computation of \eqref{eq:Dtauf1} requires the storage of the full history in $\Delta(\tau V)$.
\paragraph{3} Compute the Fourier coefficients $D_p^*$ and $\dot{D}_p^*$ of the first estimates of slip and slip rate perturbation profiles at time step $t_{n+1}$, where subscripts $p$ indicate wavenumber indices. This operation is performed using the FFT algorithm. Then estimate the stress transfer functional in the Fourier domain as (see Equation \eqref{eq:Fp})
\begin{linenomath}
  \begin{equation}\label{eq:F*}
    F^*_k = \frac{\mu|k_p|}{2}\left(-D_p^* + \sum_{k=1}^{n+1} W_p^{n-k}\dot{D}_p^k\Delta t\right),
  \end{equation}
\end{linenomath}
where $W_p^k = W(|k_p|c_\mathrm{s}t_k)$ (see Appendix \ref{ax:func}). Using an inverse FFT, compute an estimate $\phi^*_i$ of the stress transfer functional.
\paragraph{4} Compute the total strength $\tau^*_{\mathrm{f},i}=\tau_{\mathrm{f,ss},i}+\Delta\tau^*_{\mathrm{f},i}$, and the total stress $\tau_{\mathrm{stuck},i}$ that would be applied if slip rate was zero, given by
\begin{linenomath}
  \begin{equation}
    \tau_{\mathrm{stuck},i} = \tau_{\mathrm{ss},i}+\Delta\tau_\mathrm{b}(x_i)+\phi^*_i+\frac{\mu}{2c_\mathrm{s}}V_{\mathrm{ss},i}.
  \end{equation}
\end{linenomath}
Slip rate is nonzero where $\tau^*_{\mathrm{f},i}>\tau_{\mathrm{stuck},i}$. At those nodes, assign $\Delta\tau^*_i = \Delta\tau_{\mathrm{f},i}^*$, and compute the slip rate perturbation as
\begin{linenomath}
  \begin{equation}
    \Delta V_i^* = \frac{\Delta\tau_b(x_i)+\phi^*_i-\Delta\tau_{\mathrm{f},i}^*}{\mu/(2c_\mathrm{s})}.
  \end{equation}
\end{linenomath}
Where $\tau^*_{\mathrm{f},i}<\tau_{\mathrm{stuck},i}$, assign $\Delta\tau^*_i = \tau_{\mathrm{stuck},i}-\tau_{\mathrm{ss},i}$ and $\Delta V_i^*=-V_{\mathrm{ss},i}^{n+1}$.
\paragraph{5} Repeat steps 1 to 4 using $(\Delta V_i^* + \Delta V_i^n)/2$ and $(\Delta \tau_{\mathrm{f},i}^* + \Delta\tau_{\mathrm{f},i}^n)/2$ instead of $\Delta V_i^n$ and $\Delta\tau_{\mathrm{f},i}^n$, respectively. The convolutions in Equations \eqref{eq:Dtauf1} and \eqref{eq:F*}, which are the most computationally intensive steps, are not recomputed entirely but simply updated because only the last term has changed. The resulting slip, slip rate, stress and strength perturbations are the final predictions at the next time step $\Delta\delta_i^{n+1}$, $\Delta V_i^{n+1}$, $\Delta\tau_{\mathrm{f},i}^{n+1}$ and $\Delta\tau_i^{n+1}$, respectively.


\section{Perturbation amplitude}
\label{ax:ampl}

\begin{figure}[b]
  \centering
  \setfigurenum{\thesection.1}
  \includegraphics{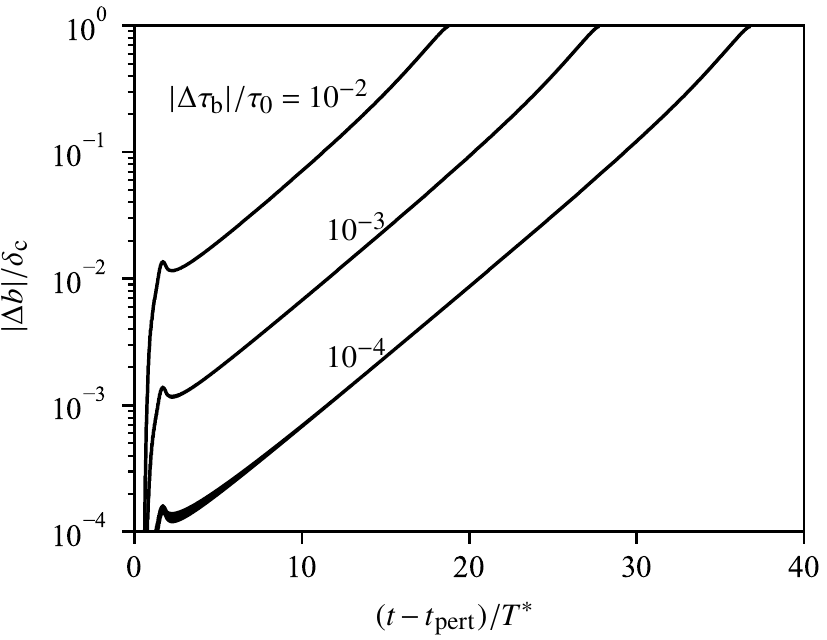}
  \caption{Time evolution of the slip perturbation for a range of amplitudes for the background stress perturbation. The initial background stress is $\tau_\mathrm{b}/\tau_0=0.7$ and the sign of the perturbation is negative.}
  \label{fig:slip_pert_amplitude}
\end{figure}

Figure \ref{fig:slip_pert_amplitude} shows the time evolution of slip perturbations following negative perturbations in background stress of $10^{-4}$, $10^{-3}$ and $10^{-2}$ in amplitude. In all simulations the reference background stress is $\tau_\mathrm{b}/\tau_0=0.7$ and the diffusivity ratio is $\alpha_\mathrm{hy}/\alpha_\mathrm{th}=1$. The growth of the slip perturbation is exponential, and the growth rate does not depend on the amplitude of the stress perturbation. The initial jump in normalised slip (and other normalised variables, see Figure \ref{fig:growthrate1}) is directly proportional to the amplitude of the stress perturbation.

\section{Expression for $\phi_{H}(x,t)$}
\label{ax:phi}

The contribution $\phi_{H}$ to the stress-transfer functional $\phi$
(Equation \ref{eq:phi}) from the second term in the expression for
the slip gradient (Equation \ref{eq:gradient}) can be written, after
moving $\partial/\partial x$ under the integral and substituting
$dx'=(d\theta(x')/dx')^{-1}d\theta$, in the following form:
\begin{linenomath}
  \begin{equation}
    \phi_{H}(x,t)=\frac{\mu}{2\pi c_\mathrm{s}}\int_{-\infty}^{t}\frac{dt'}{(t-t')^{2}}\int_{-\infty}^{t'}M'\left(\frac{x-\xi(\theta)}{c_\mathrm{s}(t-t')}\right)\frac{db}{d\theta}d\theta,
  \end{equation}
\end{linenomath}
where $M'(u)=dM/du$. Changing the order of integration in the above
double integral
\begin{linenomath}
  \begin{equation}
    \int_{-\infty}^{t}dt'\int_{-\infty}^{t'}d\theta=\int_{-\infty}^{t}d\theta\int_{\theta}^{t}dt',\label{eq:change}
  \end{equation}
\end{linenomath}
and carrying out the integral in $t'$, one finds a single-integral
expression for $\phi_{H}(x,t)$. This expression, after changing the
integration variable symbol from $\theta$ for $t'$, is given in
the main text (Equation \ref{eq:phiH} with \ref{eq:ubar}). 

\section{Deduction of expression \eqref{eq:phiD''} for $\phi_\mathrm{Dirac}(x,t)$
  based on the work of Ni and Markenscoff {[}2009{]}}

\label{ax:phiD}

Equation \eqref{eq:phiD''} can be established by accounting for the time-dependence of slip $b(t)$ in the derivation of the results of \citet{ni09} (their Equation (5.18)), who only considered dislocations with constant slip $b$. In practice, our Equation \eqref{eq:phiD''} results from carrying out the time-derivative of $b(t)$ from \citeauthor{ni09}'s Equation (3.15) to obtain a more general form of their Equation (3.16), and then equating their Equation (5.16) to the modified Equation (3.16).

It appears that \citet{ni09}, and other references of Markenscoff and co-workers give different sign (minus) in front of the logarithmic term compared to the result used here in \eqref{eq:phiD''}. A negative sign in front of the logarithmic term is inconsistent with \citeauthor{eshelby53}'s \citeyear{eshelby53} example (uniform acceleration), although one should be cautioned that there is a typographical error in the definition of $s_{0}$ used to evaluate the stress expansion (his Equation (15)) in \citeauthor{eshelby53}'s paper. It should read $s_{0}=|x_{0}-\xi(t)|/c$ instead of $s_{0}=\sqrt{t^{2}-[x-\xi(t)]^{2}}$ given directly under his Equation (15). This definition of $s_{0}$ has to be corrected in order to evaluate the logarithmic
term in the stress-expansion correctly. Markenscoff's negative sign in front of the logarithmic term is also inconsistent
with examples of direct numerical evaluation of $\phi_\mathrm{Dirac}$ for
prescribed dislocation evolution functions $\xi(t)$ and $b(t)$ (computations performed using Mathematica). Furthermore, we explicitly derive the $\ln$-term in the case of constant $v_\mathrm{r}$
and $b=b(t)$ below (Appendix \ref{ax:alternative}), which supports the sign used here. On the basis of the arguments above, we conclude that there
is a typographical sign error in the work of Markenscoff and co-workers.

\section{Alternative derivation of $\phi_\mathrm{Dirac}(x,t)$ for particular case
of a steady motion of dislocation  of variable
strength $b=b(t)$}

\label{ax:alternative}

Introducing the coordinate $X$ moving with the crack tip $X=\xi(t)-x$, and using the following notation $\Delta t=t-t'$ and $\Delta\xi=\xi(t)-\xi(t')$, we rewrite $\phi_\mathrm{Dirac}$ as
\begin{linenomath}
  \begin{equation}
    \phi_\mathrm{Dirac}(x,t)=-\frac{\mu}{2\pi c_\mathrm{s}}\int_{0}^{\infty}\frac{dM\left(\bar{u}\right)}{d\bar{u}}\frac{b(t-\Delta t)\,d\Delta t}{\Delta t^{2}},\label{eq:phiD1}
  \end{equation}
\end{linenomath}
with
\begin{linenomath}
  \begin{equation}
    \bar{u}=-\frac{X}{c_\mathrm{s}\Delta t}+\frac{\Delta\xi}{c_\mathrm{s}\Delta t}.\label{eq:ubar-1}
  \end{equation}
\end{linenomath}
The integral in \eqref{eq:phiD1} can be decomposed in that over $\Delta t$ within and outside a $t_{\epsilon}$-window. In the former, the dynamic
quantities (dislocation and rupture velocity) can be approximated
using their current rates, i.e.,
\begin{linenomath}
  \begin{equation}
    b(t-\Delta t)\approx b-\dot{b}\Delta t\qquad\frac{\Delta\xi}{\Delta t}\approx\dot{\xi}-\frac{\ddot{\xi}}{2}\Delta t\qquad(\Delta t<t_\epsilon).\label{eq:approx}
  \end{equation}
\end{linenomath}
The part of the integral \eqref{eq:phiD1} for $\Delta t>t_{\epsilon}$
is bounded by $O(b/t_{\epsilon})$. 

Thus, we focus on the integral for $\Delta t<t_{\epsilon}$ with expectation that it provides the singular part of the near-field ($X\rightarrow0$) of $\phi_{Dirac}$. Furthermore, let us restrict the consideration to \emph{steady dislocation motion} $\ddot{\xi}=0$, which simplifies the variable $\bar{u}$ dependence on $\Delta t$ to the following
\begin{linenomath}
  \begin{equation}
    \bar{u}=\frac{1}{c_\mathrm{s}}\left(v_\mathrm{r}-\frac{X}{\Delta t}\right),
  \end{equation}
\end{linenomath}
where we have renamed $\dot{\xi}=v_\mathrm{r}$)

With the above, and changing integration variable to $\bar{u}$, i.e.,
$\Delta t=\Delta t(\bar{u})$, we have
\begin{linenomath}
  \begin{equation}
    \phi_\mathrm{Dirac}^\mathrm{(singular)}(X)=-\frac{\mu}{2\pi c_\mathrm{s}}\int_{1}^{\bar{u}_{\epsilon}}\left(-\frac{\bar{u}}{\sqrt{1-\bar{u}^{2}}}\right)\left(\frac{b}{\Delta t^{2}}-\frac{\dot{b}}{\Delta t}\right)\frac{d\Delta t}{d\bar{u}}d\bar{u}
  \end{equation}
\end{linenomath}
and the ``lower'' bound of integration corresponds to minimum value
of $\Delta t$ given by $\Delta t_{1}=-X/(c_{s}-\dot{\xi})$ for which
the integrand is non-zero, i.e., $\bar{u}(\Delta t_{1})=1$, while
the ``upper'' bound $\bar{u}_{\epsilon}=\bar{u}(t_{\epsilon})$.
Explicit integration, expanding in series in small $X$ and retaining
the singular ($1/X$ and $\ln |X|$) terms leads to
\begin{linenomath}
  \begin{equation}
    \phi_\mathrm{Dirac}^\mathrm{(singular)}(X)=-\frac{\mu\sqrt{1-v_\mathrm{r}^{2}/c_\mathrm{s}^{2}}}{2\pi}\frac{b}{X}+\frac{\mu}{2\pi c_\mathrm{s}}\frac{v_\mathrm{r}/c_\mathrm{s}}{\sqrt{1-v_\mathrm{r}^{2}/c_\mathrm{s}^{2}}}\dot{b}\ln |X|.
\end{equation}
\end{linenomath}
The first in the above corresponds to the steady dislocation with
constant strength, while the second is the correction for variable
strength.


\begin{thebibliography}{47}
\providecommand{\natexlab}[1]{#1}
\expandafter\ifx\csname urlstyle\endcsname\relax
  \providecommand{\doi}[1]{doi:\discretionary{}{}{}#1}\else
  \providecommand{\doi}{doi:\discretionary{}{}{}\begingroup
  \urlstyle{rm}\Url}\fi

\bibitem[{\textit{Beeler and Tullis}(1996)}]{beeler96}
Beeler, N.~M., and T.~E. Tullis, Self-healing slip pulses in dynamic rupture
  models due to velocity-dependent strength, \textit{Bull. Seism. Soc. Am.},
  \textit{86}(4), 1130--1148, 1996.

\bibitem[{\textit{Beroza and Ellsworth}(1996)}]{beroza96}
Beroza, G.~C., and W.~L. Ellsworth, Properties of the seismic nucleation phase,
  \textit{Tectonophysics}, \textit{261}, 209--227, 1996.

\bibitem[{\textit{Brener et~al.}(2018)\textit{Brener, Aldam, Barras, Molinari,
  and Bouchbinder}}]{brener18}
Brener, E.~A., M.~Aldam, F.~Barras, J.-F. Molinari, and E.~Bouchbinder,
  Unstable slip pulses and earthquake nucleation as a nonequilibrium
  first-order phase transition, \textit{Phys. Res. Lett.}, \textit{121}(23),
  234,302, 2018.

\bibitem[{\textit{Broberg}(1978)}]{broberg78}
Broberg, K., On transient sliding motion, \textit{Geophys. J. R. astr. Soc.},
  \textit{52}, 397--432, 1978.

\bibitem[{\textit{Callias and Markenscoff}(1988)}]{callias88}
Callias, L., and X.~Markenscoff, Singular asymptotics of integrals and the
  near-field radiated from nonuniformly moving dislocations, \textit{Arch.
  Ration. Mech. Anal.}, \textit{102}(3), 273--285, 1988.

\bibitem[{\textit{Carlson and Langer}(1989)}]{carlson89}
Carlson, J.~M., and J.~S. Langer, Mechanical model of an earthquake fault,
  \textit{Phys. Rev. A}, \textit{40}(11), 6470--6484, 1989.

\bibitem[{\textit{Cochard and Madariaga}(1994)}]{cochard94}
Cochard, A., and R.~Madariaga, Dynamic faulting under rate-dependent friction,
  \textit{Pure Appl. Geophys.}, \textit{142}(3/4), 419--445, 1994.

\bibitem[{\textit{Cochard and Madariaga}(1996)}]{cochard96}
Cochard, A., and R.~Madariaga, Complexity of seismicity due to highly
  rate-dependent friction, \textit{J. Geophys. Res.}, \textit{101}(B11),
  25,321--25,336, 1996.

\bibitem[{\textit{Cochard and Rice}(1997)}]{cochard97}
Cochard, A., and J.~R. Rice, A spectral method for numerical elastodynamic
  fracture analysis without spatial replication of the rupture event,
  \textit{J. Mech. Phys. Solids}, \textit{45}(8), 1396--1418, 1997.

\bibitem[{\textit{Day et~al.}(1998)\textit{Day, Yu, and Wald}}]{day98}
Day, S.~M., G.~Yu, and D.~J. Wald, Dynamic stress changes during earthquake
  rupture, \textit{Bull. Seism. Soc. Am.}, \textit{88}(2), 512--522, 1998.

\bibitem[{\textit{{Di Toro} et~al.}(2011)\textit{{Di Toro}, Han, Hirose, {De
  Paola}, Nielsen, Mizoguchi, Ferri, Cocco, and Shimamoto}}]{ditoro11}
{Di Toro}, G., R.~Han, T.~Hirose, N.~{De Paola}, S.~Nielsen, K.~Mizoguchi,
  F.~Ferri, M.~Cocco, and T.~Shimamoto, Fault lubrication during earthquakes,
  \textit{Nature}, \textit{471}, 494--498, 2011.

\bibitem[{\textit{Dunham and Archuleta}(2005)}]{dunham05}
Dunham, E.~M., and R.~J. Archuleta, Near-source ground motion from steady state
  dynamic rupture pulses, \textit{Geophys. Res. Lett.}, \textit{32}, L03302,
  \doi{10.1029/2004GL021793}, 2005.

\bibitem[{\textit{Elbanna and Heaton}(2012)}]{elbanna12}
Elbanna, A.~E., and T.~H. Heaton, A new paradigm for simulating pulse-like
  ruptures: the pulse energy equation, \textit{Geophys. J. Int.}, \textit{189},
  1797--1806, 2012.

\bibitem[{\textit{Eshelby}(1953)}]{eshelby53}
Eshelby, J.~D., The equation of motion of a dislocation, \textit{Phys. Rev.},
  \textit{90}(2), 248--255, 1953.

\bibitem[{\textit{Freund}(1979)}]{freund79}
Freund, L.~B., The mechanics of dynamic shear crack propagation, \textit{J.
  Geophys. Res.}, \textit{84}, 2199--2209, 1979.

\bibitem[{\textit{Gabriel et~al.}(2012)\textit{Gabriel, Ampuero, Dalguer, and
  Mai}}]{gabriel12}
Gabriel, A.-A., J.-P. Ampuero, L.~A. Dalguer, and P.~M. Mai, The transition of
  dynamic rupture styles in elastic media under velocity-weakening friction,
  \textit{J. Geophys. Res.}, \textit{117}, B09311, \doi{10.1029/2012JB009468},
  2012.

\bibitem[{\textit{Galetzka et~al.}(2015)\textit{Galetzka, Melgar, Genrich,
  Geng, Owen, Lindsey, Xu, Bock, Avouac, Adhikari, Upreti, {Pratt-Sitaula},
  Bhattarai, Sitaula, Moore, Hudnut, Szeliga, Normandeau, Fend, Flouzat,
  Bollinger, Shrestha, Koirala, Gautam, Bhatterai, Gupta, Kandel, Timsina,
  Sapkota, Rajaure, and Maharjan}}]{galetzka15}
Galetzka, J., D.~Melgar, J.~F. Genrich, J.~Geng, S.~Owen, E.~O. Lindsey, X.~Xu,
  Y.~Bock, J.-P. Avouac, L.~B. Adhikari, B.~N. Upreti, B.~{Pratt-Sitaula},
  T.~N. Bhattarai, B.~P. Sitaula, A.~Moore, K.~W. Hudnut, W.~Szeliga,
  J.~Normandeau, M.~Fend, M.~Flouzat, L.~Bollinger, P.~Shrestha, B.~Koirala,
  U.~Gautam, M.~Bhatterai, R.~Gupta, T.~Kandel, C.~Timsina, S.~N. Sapkota,
  S.~Rajaure, and N.~Maharjan, Slip pulse and resonance of the {K}athmandu
  basin during the 2005 {G}orkha earthquake, {N}epal, \textit{Science},
  \textit{349}(6252), 1091--1095, 2015.

\bibitem[{\textit{Garagash}(2012)}]{garagash12}
Garagash, D.~I., Seismic and aseismic slip pulses driven by thermal
  pressurization of pore fluid, \textit{J. Geophys. Res.}, \textit{117},
  B04314, \doi{doi:10.1029/2011JB008889}, 2012.

\bibitem[{\textit{Heaton}(1990)}]{heaton90}
Heaton, T.~H., Evidence for and implications of self-healing pulses of slip in
  earthquake rupture, \textit{Phys. Earth Planet. Int.}, \textit{64}, 1--20,
  1990.

\bibitem[{\textit{Johnson}(1990)}]{johnson90}
Johnson, E., On the initiation of unidirectional slip, \textit{Geophys. J.
  Int.}, \textit{101}, 125--132, 1990.

\bibitem[{\textit{Johnson}(1992)}]{johnson92e}
Johnson, E., The influence of the lithospheric thickness on bilateral slip,
  \textit{Geophys. J. Int.}, \textit{108}, 151--160, 1992.

\bibitem[{\textit{Lapusta and Rice}(2003)}]{lapusta03}
Lapusta, N., and J.~R. Rice, Low-heat and low-stress fault operation in
  earthquake models of statically strong but dynamically weak faults,
  \textit{Eos. Trans. AGU}, \textit{84}(46), {Fall} Meet. Suppl., Abstract
  S51B-02, 2003.

\bibitem[{\textit{Lapusta et~al.}(2000)\textit{Lapusta, Rice, {Ben-Zion}, and
  Zheng}}]{lapusta00}
Lapusta, N., J.~R. Rice, Y.~{Ben-Zion}, and G.~Zheng, Elastodynamic analysis
  for slow tectonic loading with spontaneous rputure episodes on faults with
  rate-and-state-dependent friction, \textit{J. Geophys. Res.},
  \textit{105}(B10), 23,765--23,789, 2000.

\bibitem[{\textit{Markenscoff}(1980)}]{markenscoff80}
Markenscoff, X., The transient motion of a nonuniformly moving dislocation,
  \textit{J. Elast.}, \textit{10}(2), 193--201, 1980.

\bibitem[{\textit{Ni and Markenscoff}(2008)}]{ni08}
Ni, L., and X.~Markenscoff, The self-force and effective mass of a generally
  accelerating dislocation {I}: {S}crew dislocation, \textit{J. Mech. Phys.
  Solids}, \textit{56}, 1348--1379, \doi{10.1016/j.jmps.2007.09.002}, 2008.

\bibitem[{\textit{Ni and Markenscoff}(2009)}]{ni09}
Ni, L., and X.~Markenscoff, The logarithmic singularity of a generally
  accelerating dislocation from the dynamic energy-momentum tensor,
  \textit{Math. Mech. Solids}, \textit{14}, 38--51, \doi{10.1177/
  1081286508092601}, 2009.

\bibitem[{\textit{Noda and Lapusta}(2010)}]{noda10}
Noda, H., and N.~Lapusta, Three-dimensional earthquake sequence simulations
  with evolving temperature and pore pressure due to shear heating: Effect of
  heterogeneous hydraulic diffusivity, \textit{J. Geophys. Res.}, \textit{115},
  B12314, \doi{10.1029/2010JB007780}, 2010.

\bibitem[{\textit{Noda et~al.}(2009)\textit{Noda, Dunham, and Rice}}]{noda09}
Noda, H., E.~M. Dunham, and J.~R. Rice, Earthquake ruptures with thermal
  weakening and the operation of major faults at low overall stress levels,
  \textit{J. Geophys. Res.}, \textit{114}, B07302, \doi{10.1029/2008JB006143},
  2009.

\bibitem[{\textit{Olsen et~al.}(1997)\textit{Olsen, Madariaga, and
  Archuleta}}]{olsen97}
Olsen, K.~M., R.~Madariaga, and R.~J. Archuleta, Three-dimensional dynamic
  simulation of the 1992 {L}anders earthquake, \textit{Science},
  \textit{278}(5339), 834--838, 1997.

\bibitem[{\textit{Pellegrini}(2010)}]{pellegrini10}
Pellegrini, Y.-P., Dynamic peierls-nabarro equations for elastically isotropic
  crystals, \textit{Phys. Rev. B}, \textit{81}, 024,101,
  \doi{10.1103/PhysRevB.81.024101}, 2010.

\bibitem[{\textit{Perrin et~al.}(1995)\textit{Perrin, Rice, and
  Zheng}}]{perrin95}
Perrin, G., J.~R. Rice, and G.~Zheng, Self-healing pulse on a frictional
  interface, \textit{J. Mech. Phys. Solids}, \textit{43}(9), 1461--1495, 1995.

\bibitem[{\textit{Platt et~al.}(2014)\textit{Platt, Rudnicki, and
  Rice}}]{platt14}
Platt, J.~D., J.~W. Rudnicki, and J.~R. Rice, Stability and localization of
  rapid shear in fluid-saturated fault gouge, 2. {L}ocalized zone width and
  strength evolution, \textit{J. Geophys. Res.}, \textit{119}, 4334--4359,
  \doi{10.1002/2013JB010711}, 2014.

\bibitem[{\textit{Platt et~al.}(2015)\textit{Platt, Viesca, and
  Garagash}}]{platt15b}
Platt, J.~D., R.~C. Viesca, and D.~I. Garagash, Steadily propagating slip
  pulsees driven by thermal decomposition, \textit{J. Geophys. Res.},
  \textit{120}, 6558--6591, \doi{10.1002/2015JB012200}, 2015.

\bibitem[{\textit{Rice}(1980)}]{rice80}
Rice, J.~R., The mechanics of earthquake rupture, in \textit{Physics of the
  {E}arth's Interior}, edited by A.~M. Dziewonski and E.~Boschi, Proc. Intl.
  School of Physics E. Fermi, Italian Physical Society/North Holland Publ. Co.,
  1980.

\bibitem[{\textit{Rice}(2006)}]{rice06}
Rice, J.~R., Heating and weakening of faults during earthquake slip, \textit{J.
  Geophys. Res.}, \textit{111}, B05311, \doi{10.1029/2005JB004006}, 2006.

\bibitem[{\textit{Rice et~al.}(2005)\textit{Rice, Sammis, and
  Parsons}}]{rice05}
Rice, J.~R., C.~G. Sammis, and R.~Parsons, Off-fault secondary failure induced
  by a dynamic slip pulse, \textit{Bull. Seism. Soc. Am.}, \textit{95}(1),
  109--134, 2005.

\bibitem[{\textit{Rice et~al.}(2014)\textit{Rice, Rudnicki, and
  Platt}}]{rice14}
Rice, J.~R., J.~W. Rudnicki, and J.~D. Platt, Stability and localization of
  rapid shear in fluid-saturated fault gouge, 1. {L}inearized stability
  analysis, \textit{J. Geophys. Res.}, \textit{119}, 4311--4333,
  \doi{10.1002/2013JB01071}, 2014.

\bibitem[{\textit{Rosakis}(2001)}]{rosakis01}
Rosakis, P., Supersonic dislocation kinetics from an augmented {P}eierls model,
  \textit{Phys. Res. Lett.}, \textit{86}(1), 95--98, 2001.

\bibitem[{\textit{Rubinstein et~al.}(2004)\textit{Rubinstein, Cohen, and
  Fineberg}}]{rubinstein04}
Rubinstein, S.~M., G.~Cohen, and J.~Fineberg, Detachment fronts and the onset
  of dynamic friction, \textit{Nature}, \textit{430}, 1005--1009, 2004.

\bibitem[{\textit{Schmitt et~al.}(2011)\textit{Schmitt, Segall, and
  Matsuzawa}}]{schmitt11}
Schmitt, S.~V., P.~Segall, and T.~Matsuzawa, Shear heating-induced thermal
  pressurization during earthquake nucleation, \textit{J. Geophys. Res.},
  \textit{116}, B06308, \doi{10.1029/2010JB008035}, 2011.

\bibitem[{\textit{Trefethen}(2000)}]{trefethen00}
Trefethen, L.~N., \textit{Spectral methods in \textsc{{M}atlab}}, vol.~10,
  Society for Industrial Mathematics, 2000.

\bibitem[{\textit{Viesca and Garagash}(2015)}]{viesca15}
Viesca, R.~C., and D.~I. Garagash, Ubiquitous weakening of faults due to
  thermal pressurization, \textit{Nat. Geosci.}, \doi{10.1038/ngeo2554}, 2015.

\bibitem[{\textit{Viesca and Garagash}(2018)}]{viesca18}
Viesca, R.~C., and D.~I. Garagash, Numerical methods for coupled fracture
  problems, \textit{J. Mech. Phys. Solids}, \textit{113}, 13--34, 2018.

\bibitem[{\textit{Wald and Heaton}(1994)}]{wald94}
Wald, D.~J., and T.~H. Heaton, Spatial and temporal distribution of slip for
  the 1992 {L}anders, {C}alifornia, earthquake, \textit{Bull. Seism. Soc. Am.},
  \textit{84}(3), 668--691, 1994.

\bibitem[{\textit{Weertman}(1969)}]{weertman69}
Weertman, J., Dislocation motion on an interface with friction that is
  dependent on sliding velocity, \textit{J. Geophys. Res.}, \textit{74}(27),
  6617--6622, 1969.

\bibitem[{\textit{Weertman}(1980)}]{weertman80}
Weertman, J., Unstable slippage across a fault that separates elastic media of
  different elastic constants, \textit{J. Geophys. Res.}, \textit{85}(B8),
  1455--1461, 1980.

\bibitem[{\textit{Zheng and Rice}(1998)}]{zheng98}
Zheng, G., and J.~R. Rice, Conditions under which velocity-weakening friction
  allows a self-healing versus a cracklike more of rupture, \textit{Bull.
  Seism. Soc. Am.}, \textit{88}(6), 1466--1483, 1998.

\end{thebibliography}

\end{article}

\end{document}